\documentclass[11pt,twoside]{article}
\usepackage[utf8]{inputenc}
\usepackage{amsmath}
\usepackage{amssymb}
\usepackage[numbers]{natbib} 
\usepackage{graphics}
\usepackage{latexsym}
\usepackage{amsfonts}
\usepackage{graphicx}
\usepackage{caption}
\usepackage{subfig}
\usepackage{pstricks}
\usepackage{epigraph}
\usepackage{picture}
\usepackage[english]{babel}
\usepackage{enumitem}

\author{
Alessio E.\ Biondo$^a$\thanks{\textit{Corresponding author}. Email:
ae.biondo@unict.it. \newline $^a$Department of Economics and
Business, University of Catania, Corso Italia 55, 95129 Catania,
Italy.
\newline $^b$Department of Physics and Astronomy, University of Catania, Via S.\ Sofia 64, 95123 Catania, Italy.
\newline $^c$INFN Section of Catania, Via S.\ Sofia 64, 95123 Catania, Italy.} ,
Alfio Giarlotta$^a$,
Alessandro Pluchino$^{bc}$,
Andrea Rapisarda$^{bc}$ }
\date{}

\title{\bf Perfect Information vs Random Investigation: \\
Safety Guidelines for a Consumer in the Jungle of Product Differentiation}

\def\Sat{\textsf{Sat}}
\def\Kn{\textsf{Kn}}
\def\Lev{\textsf{Lev}}
\def\Aw{\textsf{Aw}}

\begin{document}

\maketitle

\begin{abstract}
\noindent We present a graph-theoretic model of consumer choice, where final
decisions are shown to be influenced by information and knowledge,
in the form of individual awareness, discriminating ability, and
perception of market structure. Building upon the distance-based
Hotelling's differentiation idea, we describe the behavioral
experience of several prototypes of consu.pdfmers, who walk a
hypothetical cognitive path in an attempt to maximize their
satisfaction. Our simulations show that even consumers endowed with a
small amount of information and knowledge may reach a very high
level of utility. On the other hand, complete ignorance negatively
affects the whole consumption process. In addition, rather unexpectedly,
a random walk on the graph reveals to be a winning strategy, below a minimal threshold of information and knowledge.

\bigskip
\noindent \textbf{Keywords:} Product differentiation; Hotelling; consumer behavior; information; knowledge; random choice.
\end{abstract}

\section{Introduction}

Since the seminal work of Chamberlin \cite{chamberlin1962}, the
economics literature on monopolistically competitive markets has
been rooted on the idea of product differentiation
\cite{dixitetal1977}: goods are provided by many producers in
several versions and models, despite they aim at satisfying the very
same need. Such a market framework can be easily experienced
everyday in real-life markets, where several products compete with
each other in offering very similar services to the final consumer.

Product differentiation has been studied from two perspectives: (i)
\emph{horizontal}, often referred to as spatial
differentiation \cite{hotelling1929,salopsteven1979}, and (ii)
\emph{vertical} \cite{gabszewicz1979, gabszewicz-thisse80,shaked1982relaxing, shakedsutton1983}. If a good is defined as a bundle
of characteristics (in the sense of Lancaster \cite{lancaster1966}),
the distinction between horizontal and vertical differentiation is
that the former refers to goods with different features and the same
qualitative level, whereas the latter takes into account differences
in quality. However, as Cremer and Thisse have shown \cite{cremerthisse1991},
results of Hotelling-type models and vertical differentiation models
are basically equivalent. In this paper we assume that the
differentiation has a comprehensive characterization: in fact,
different features and qualitative levels will be described in a
unique multidimensional environment.\footnote{In other words, we
refer to a higher-order differentiation concept, which embeds both
the number of product features and their intensity. More precisely,
we implicitly take a multi-criteria approach, in which every feature
(including price) is suitably modeled and weighted: the reader can refer to literature about the \emph{multi-attribute utility theory} and the
\emph{outranking} approach, \cite{keeneyetal1993} \cite{roy1985}.}

Consumer's behavioral choice is studied in the literature from
several, yet complementary, points of view. From a buyer's
perspective, differentiation among goods is advantageous insofar as
it helps in seeking a feasible good offering the highest
satisfaction. In this sense, an increase in differentiation is
usually associated to a positive value for buyers, because it
increases the possibility of a making a selection that is close to
an ideal target.\footnote{For a detailed analysis of the possible
reasons why consumers seek variety, we refer the reader
to \cite{kahn1995consumer}.} Monopolistic competition and oligopoly characterize the
largest part of the real market experience. These two market structures share
many common characteristics; however, they differ from each other in the way the total
market output is looked at. In fact, oligopolistic firms explicitly consider the effect of their role
in determining the total market output, whereas monopolistically competitive firms
consider the aggregate market output as exogenously given in the process of setting their
individual production, \cite{cellinietal2015}. We focus our attention on the demand side of a
\emph{monopolistically competitive market}, where firms' supply
composition is taken as an exogenous environmental configuration.
Our goal is to study a consumer's individual choice process, that
is, to analyze the way in which she  selects a final product among
several version of it supplied by different brands. To this aim, we
make the (non-simplifying) assumption that the consumer considers
quality, features, and prices as a unified multi-criteria
discrimination concept.

Market failure due to asymmetric information occurs when the buyer and the seller have two
different informative sets and the more aware of them tries to extract
a supplementary benefit at the expenses of the other \cite{akerlof1970market,
spence1976, rileyJohnetal2001} . Indeed, the case of a consumer who does not know the
 complete characterization of the goods she needs must be considered: that consumer
chooses blinded by her ignorance \cite{biondo2014organic}. The result is a
lower level of satisfaction than the potential one. This may require a policy
intervention to solve the resulting inefficiency. To further support this point of view, consider
the perverse effect of differentiation: the ``tyranny of choice" -- a term coined by Schwartz
\cite{schwartzbarry2004} -- represents the situation when the abundance of
available choices on the market may even become undesirable if the
consumer cannot count on a valid consciousness to guide her
selection. In this respect, some authors stress the role of advertisement in informing 
consumers \cite{grossmanetal1984}. However, adverts may barely provide a neutral source 
of knowledge. Thus, the model here proposed will show effects of advertising.

The relevance of \emph{information} and \emph{knowledge} in
consumption is widely recognized, because of their ability to
influence the consumer's attitude in perception and searching.
Indeed, some analysis supports the point of view that a lack of
knowledge may truly generate a reduction of the consumer's judgement
capacity \cite{sirieixetal2013}. In our opinion this conclusion cannot be fully
ascribed to the well-known ``bounded rationality" concept, which
appears more linked to the completeness of the informative set, as
advocated in \cite{simon1955behavioral, simon1976substantive} and in \cite{selten1978chain}. In this paper we explicitly emphasize the semantic difference
between the two concepts of ``information" and ``knowledge". Our
interpretation of these two terms relies on a suggested distinction
between (i) what the consumer knows with regards to the current
transaction being concluded (which should be considered as
``information'', as generally accepted in literature), and (ii) what
the consumer knows in a broader sense, in the form of cultural
background and capability to understand and to evaluate (which is a
notion of ``knowledge''). The interested reader may refer to existing surveys
\cite{lipman1995information, rubinstein1998modeling}. A possible rationale for the reduction of the consumer's judgement capacity can be ascribed to the complexity of choice environment \cite{swait2001choice}. In fact, it is known that consumers decide to make purchases for many reasons, which range from very basic
needs to utmost volatile instincts. As a consequence, even their perceived
desires/needs and the way they are expressed respond to a very wide
set of stimuli, as documented in an ample literature \cite{balasubramanian2005consumers, sivaramakrishnan2007giving, chartrandetal2008}. Also, there is evidence that choices made
in consumption may affect the development dynamics of the economy \cite{zaccariacristellitacchellapietronero}.

We approach the problem of individual consumer's choice by
describing the different phases of the informative process that
leads the buyer to her final selection. In this sense, we assume
that the consumer is able to build an ``experiential cognitive
map'', where all the variables are taken into account in a spatial
sense. Our approach basically belongs to the well-known framework of
consumer decision-making research \cite{bettman1991}, which
considers consumption as a process that generally involves the
following phases: (1) problem recognition (i.e., market selection
according to the need being satiated); (2) information search; (3)
evaluation of alternatives; (4) final choice (i.e., purchase); (5)
post-purchase evaluation. We assume that the consumer has chosen the
market where to enter, and focus on phases (2), (3), and (4). We
will neglect phase (5).

In order to build a spatial configuration, here we use a
Hotelling-type approach \cite{hotelling1929, salopsteven1979}, to qualify the
satisfaction in terms of the distance from the ``perfect choice".
Our model informally follows a multi-attribute approach: we
assume that the consumer's exploration will reveal her behavioral
attitude in recognizing and appreciating desired characteristics.
Thus, the set of available options is analyzed, and, after
comparisons, the most satisfying alternative is chosen
\cite{bettmanjames1979, bettmanetal1998, 
aranaleon2009}. 

Market complexity affects many aspects of the decision process.
In fact, motivational elements and topological configurations
of attributes space unavoidably link the maximizing
choice to the effort put in the search \cite{depalmaetal94}. 
This is the reason why our model is based on two distinct
features: (i) the consumer's ability in discriminating similar
goods, and (ii) her knowledge of the market structure. Furthermore, a very strong 
influence on present choices and decisions is exerted by similar previous
experiences \cite{nosofsky2000exemplar, gilboa2001cognitive,
juslin2008emotional, broder2010cue,
scheibehenne2015different}. As a consequence, we assume that a purchase decision
passes through an analysis that takes into account every cognitive
element of the consumer's activity, either present or stored in
memory from past experiences \cite{nosofskyrobert1984}. The theoretical
context employed here allows individuals to adaptively select
strategies in order to manage the actual situation at their best \cite{scheibehenneetal2013} \cite{berkowitsch2014rigorously}. 

In this paper, we investigate the existence of some ``safety guidelines" to help consumers with different degrees of information to make their choices. Inspired by previous studies on the beneficial role of randomness in socio-economic systems, we actually test the effectiveness of several strategies to reach a final decision, ranging from the hypothetical scenario of perfect information, to a completely random walk. As it was found for financial markets \cite{noiJSP,noiPLOSONE,noiPRE,noiCP}, for career advancements in managerial organizations \cite{AAPeter1,AAPeter2}, and for efficiency of political institutions \cite{AAParl}, we anticipate that also in this case randomness gives positive results. 

The paper is organized as follows. Section 2 presents the model, in
its static and dynamic features. In Section 3 we run various types
of simulations, and present numerical results. Conclusions and future directions of research are addressed in Section 4.


\section{The model: statics and dynamics} \label{SECT:statics_and_dynamics_model}

In this section we describe our model. We first deal with
its static features, which are related to the exogenous market
structure (\ref{SUBSECT:model_for_market}), the endogenous
consumer's characteristics (\ref{SUBSECT:model_for_consumer}), and
the interaction between market and consumer
(\ref{SUBSECT:gravitational field}). Successively, we describe the
dynamical procedure employed by the consumer to explore the market, gathering information in an attempt to reach her target
(\ref{SUBSECT:dynamics_of_model}).


\subsection{The market} \label{SUBSECT:model_for_market}

The topological structure of the market is represented by a graph
with three types of nodes, identified by different shapes, see
Figure 1(a). Formally, the graph has the theoretic structure of a \emph{forest}, that is, a free union of \emph{trees} (connected acyclic
graphs).\footnote{Recall that a \emph{graph} $G$ is a pair $(X,E)$,
where $X$ is a nonempty set of \emph{nodes}, and $E$ is a (possibly
empty) set of (undirected) \emph{edges}, with an edge being a subset
of $X$ having size two. Two distinct nodes $x,y \in X$ are
\emph{adjacent} if $\{x,y\}$ is an edge in $E$. A \emph{path} in a
graph $G=(X,E)$ is a sequence $(x_1,\ldots,x_{k})$ of $k \geq 2$
adjacent nodes, in the sense that the set $\{x_i,x_{i+1}\}$ is an
edge in $E$ for each $i = 1,\ldots,k-1$. A graph is \emph{connected}
whenever any two distinct nodes are joined by a path, and is
\emph{acyclic} if there are no paths beginning and ending at the
same node. For further details, refer to \cite{bollobas}.} Each tree is star-shaped, and represents a ``cluster'':
for instance, the graph in Figure~1(a) has three clusters. The
meaning of the three types of nodes is described below.

\begin{description}
  \item[\bf Central nodes.] These nodes are located at the center of the trees composing the forest,
  and are denoted by pentagons.
  They are the \emph{hubs} (i.e., highly connected nodes) of the graph,
  and may represent either a ``brand'' or a  ``category'' of products. For example, in the market of cameras,
  we can look at each central node as a specific producer
  (Canon, Minolta, Nikon, etc.) or as a type of camera (bridge, mirrorless, reflex, etc.).
  \item[\bf Terminal nodes.] These are the \emph{leaves} of the forest
  (i.e., the nodes with degree 1), and are denoted by circles.\footnote{Recall
  that the \emph{degree} of a node $x$ is the number of nodes that are
  adjacent to $x$. The nodes adjacent to $x$ are also called the \emph{first neighbors} of $x$.}
  The terminal nodes represent all final products that are sold on
  the specific market at hand. For instance, in dealing with the market of cellular phones,
  a terminal node may be a 4-tuple of the type
  $\big\langle$Apple\,,\:iPhone6\,,\:32\:\textsf{GB}\,,\:silver$\big\rangle$.
  \item[\bf Intermediate nodes.] These are the nodes that belong to paths connecting
  the hub of a cluster to the terminal nodes of the same cluster, and are denoted by  squares.
  Intermediate nodes represent the distinct phases of a consumer's \emph{informative journey},
  intended as a cognitive walk that guides her from a brand/category to a final product present in the market.
  To illustrate the role of intermediate nodes, consider the following example in a computer market.
  A consumer decides to buy a final product represented by the $6$-tuple
  \begin{center}
  \small{$\big\langle$Apple\,,\:Macbook
  Air\,,\:$13''$,\:i7\:-\:2.2\:\textsf{GHz}\,,\:8\:\textsf{GB\:SDRAM}\,,\:512\:\textsf{GB}$\big\rangle$.}
  \end{center}
  Here the central node is the brand $\langle$Apple$\rangle$, the terminal node is a laptop with
  some specified features, and the intermediate nodes are all restrictions of the $6$-tuple
  to the first $i$ components,\footnote{Specifically, the four
  intermediate nodes are $\big\langle$Apple\,,\:Macbook Air$\big\rangle$,
  $\big\langle$Apple\,,\:Macbook Air\,,\:$13''$$\big\rangle$,
  $\big\langle$Apple\,,\:Macbook
  Air\,,\:$13''$,\:i7\:-\:2.2\:\textsf{GHz}$\big\rangle$, and
  $\big\langle$Apple\,,\:Macbook Air\,,\:$13''$,\:i7\:-\:2.2\:\textsf{GHz}\,,\:8\:\textsf{GB\:SDRAM}$\big\rangle$.}
  with $i = 2,3,4,5$. Thus, each product can be identified with
  an $n$-tuple of features specified by the sequence of $n-1$ informative
  steps needed to reach it, starting from the central node (which represents its first feature).
  In this way, a product actually sold in the market is completely determined by the topology of the cluster
  (brand/category) to which it belongs.
\end{description}
It is important to emphasize that in our approach the market
structure is totally exogenous. Therefore, at this initial stage,
the distinct clusters constituting the market are not connected to
each other, and live in a space that is \emph{not} endowed with any
metric structure.

\begin{figure}
\centering \subfloat [][Topological structure of the market]
{\includegraphics[width=0.485\textwidth]{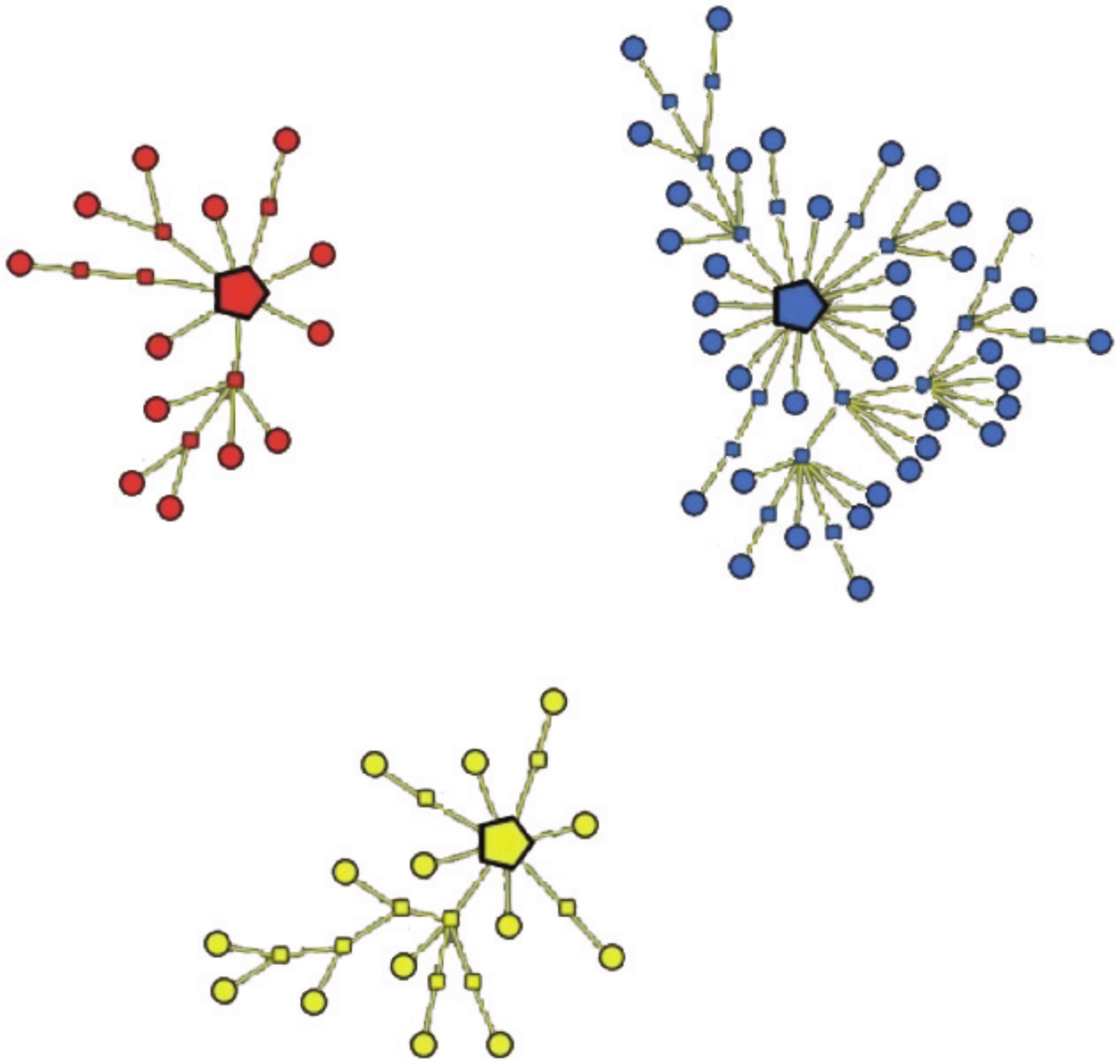}} \quad
\subfloat [][Metric structure induced by consumer's satisfaction]
{\includegraphics[width=0.485\textwidth]{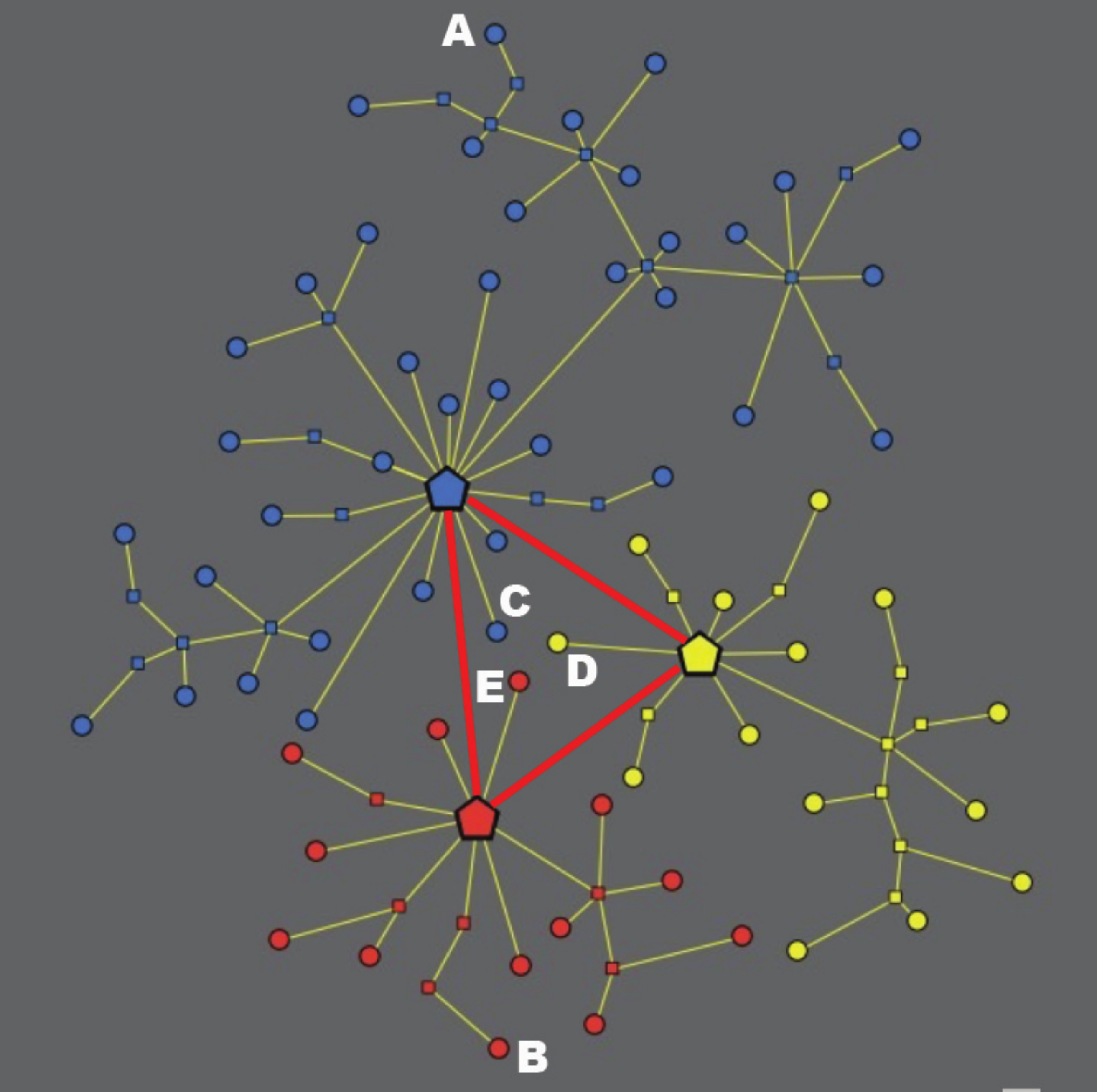}}
\caption{\footnotesize \emph{Graph representation of a market with
three clusters.} The clusters are identified by different colors.
The red edges in the right picture represent the consumer's
knowledge (which is ``complete'' in this case).}
\end{figure}


\subsection{The consumer} \label{SUBSECT:model_for_consumer}

Once that both the number and the topological structure of the
clusters are determined, the market has to be evaluated by
consumers. In fact, each (type of) consumer has different features,
preferences and perceptions, which constitute the ``lens'' through
which the market is viewed by her. Below we describe in detail how
the consumer's point of view may affect the representation of the
market, in terms of her: \emph{satisfaction}, \emph{knowledge}, and \emph{discrimination ability}.

\begin{description}

\item[\bf Satisfaction metric.]
The presence of a personal point of view naturally induces a
deformation of the graph structure, which however leaves unchanged
the original topology of the market. In fact, the market assumes a
shape that reflects the existence of an underlying
\emph{satisfaction metric}, defined in an appropriate space and
intrinsic to each (type of) consumer: see Figure 1(b), where, for
the sake of graphical representation, we embed the graph into the
two-dimensional space $(\mathbb{R}^2,d)$, with $d$ denoting the
standard Euclidean distance.

In order to better specify such an intrinsic satisfaction metric, we
assume that each consumer has in mind an ``ideal goal'', a
hypothetical target product with several well-specified
characteristics. This target, which may or may not exist in the
market, occupies a specific position in the Euclidean space, say, $P^*
\equiv (x_{P^*},y_{P^*}) \in \mathbb{R}^2$. In particular, if the target is
a real product sold in the market, then it is located on a terminal
node of the graph.

A natural assumption of our model is that a consumer purchasing her
ideal product $P^*$ will retrieve from it the maximum \emph{ex-post}
satisfaction, that is, $\Sat(P^*) =1$. On the other hand, purchasing a
product different from her target and corresponding to a terminal
node $P\equiv(x_P,y_P) \in \mathbb{R}^2$ will provide her with a
non-perfect satisfaction, that is, $0\leq \Sat(P) < 1$ in this case.
Note that the satisfaction $\Sat(P)$ retrieved from a non-ideal
product $P$ shall naturally depend on the Euclidean distance
$d(P,P^*)= \big((x_P-x_{P^*})^2 + (y_P-y_{P^*})^2\big)^{1/2}$ of the chosen
product $P$ from the target $P^*$. More generally, we define the
satisfaction $\Sat(P)$ associated to any point $P \in \mathbb{R}^2$
by
\begin{equation}
\label{satisfaction} \Sat(P) \;:= \; 1 \, - \,
\frac{d(P,P^*)}{d_{\max}}
\end{equation}
where $d_{\max}$ is the maximum possible distance in the bounded
part of the considered space $\mathbb{R}^2$ (which is a square, in
our case). Due to its generality, this definition of satisfaction
works for any pair of nodes (not necessarily terminal).

It is important to note that in our model the distance between two
terminal nodes is independent from the set of features of the
corresponding products. For instance, the top blue node $A$ and the
lower red node $B$ in Figure 1(b) -- which are very far from each
other in the underlying metric space -- might well have very similar
features; nevertheless, it is possible that the satisfaction
deriving from buying $A$ is very different from that deriving from
buying $B$, even just because they belong to different
clusters.\footnote{Just to give a concrete example, think to a
consumer whose target is an \emph{Apple iPhone} with certain particular
features: if she  buys a \emph{Samsung Galaxy} smartphone with very similar
characteristics, then it is likely that she  will be much less
satisfied, since in this specific case the importance of the brand
dominates the other features of the product.} On the other hand, the
three nodes of different colors placed at the center of the metric
space ($C$, $D$ and $E$) might present very different features
despite being very close to each other (hence giving a similar
satisfaction to the consumer).

\item[\bf Knowledge of clusters.]
In reality, consumers' knowledge of the exogenous market structure
may vary quite a lot. 
We measure knowledge in terms of perception of the existence of
clusters; those clusters that are known to the consumer are labeled
as \emph{active}. Figure 1(b) graphically describes a deformed
topological structure of a simple market with three clusters, as
viewed by a perfectly informed consumer, i.e., a consumer with a
maximum knowledge $\Kn_{\max}=1$. In the represented case, the three
clusters are all active, in fact their central nodes are connected
to each other by red edges (whereas the \emph{intra-clusters} links
are colored in yellow). Note such a perfectly informed consumer
could theoretically visit all the nodes of the graph during her
informative journey, using both red and yellow edges.

The situation is quite different for a consumer with an imperfect
knowledge $\Kn<1$. In Figure~2 we describe a market with seven
clusters, as it is viewed by a consumer with a knowledge $\Kn=0.5$:
in this case only three of the seven clusters are active, hence the
informative journey of this type of consumer will be limited to the
three corresponding brands/categories. In Figure~2 we also represent
the following additional features:
\begin{itemize}
    \item[(1)] A possible position of the consumer at the beginning of her
    journey over the graph, indicated by a red human shape located
    on a node belonging to the top-right (active) cluster. In what follows, we refer to
    this initial node as the \emph{source} of her psychological journey. (For example, in the process
    of searching for a new cellular phone to buy, the consumer might
    start her exploration from the node of the graph that represents
    -- or is very close to -- her old phone.)
    \item[(2)] A possible position of her target product, indicated by a shape with
    concentric red circles. In the represented case, the target is located on a terminal
    node of the middle-right (inactive) cluster. (However, as already emphasized,
    the location of the target may well fail to coincide with a terminal node of a cluster, thus
    indicating that the consumer's ideal product is not present in the market.)
\end{itemize}
In the described situation, despite the target coincides with a
terminal node of the graph, the consumer will never be able to
reach it, due to her imperfect knowledge: in fact, the target
belongs to an inactive cluster. Therefore, the consumer's greatest
ambition can only be to get as close as possible to her target,
while remaining within the subset of the market that is reachable by
his partial knowledge.

\begin{figure}
\centering
\includegraphics[width=0.5\textwidth]{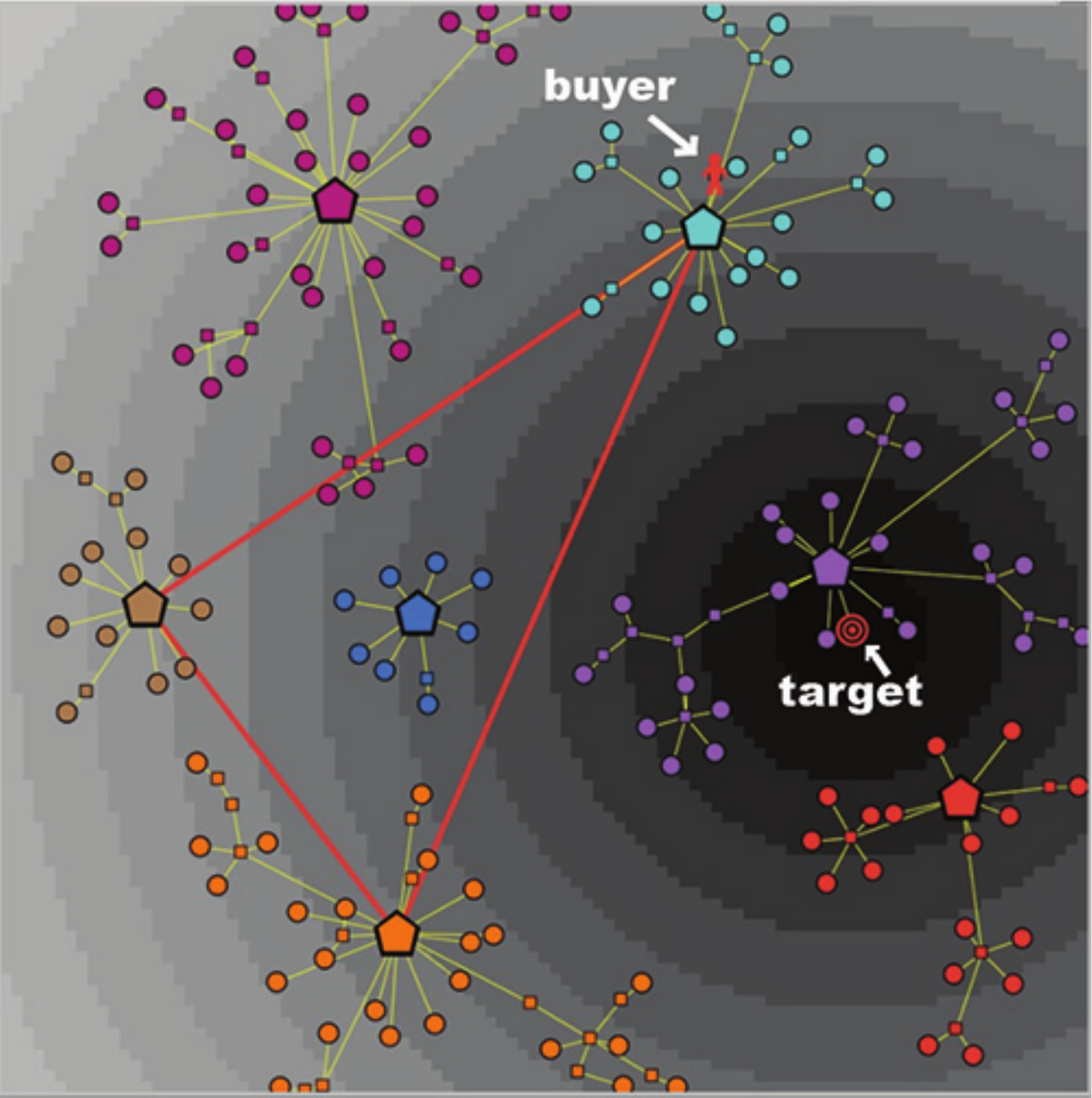}
\caption{\footnotesize \emph{A market with seven clusters, as viewed
by a consumer with moderate knowledge and high discrimination
power.} The consumer has a partial knowledge ($\Kn=0.5$) and a quite
high number of indifference levels ($N_\Lev=14$), represented by
annuli with different shades of gray. The indifference areas are
centered at the target, which is located on a node, and is denoted
by concentric red circles. The random initial position of the buyer
is located on a node of the top-right cluster, and is denoted by a
human shape.}
\end{figure}

\item[\bf Preference structure and discrimination ability.]
The last feature that influences the market evaluation by the
consumer is her preference structure. In our model, we assume that
the buyer has only a sort of ``fuzzy'' perception of the level of
satisfaction associated to each point of the graph, until he
actually reaches a node and physically explores it. Specifically, we
distinguish (1) an ``\emph{ex-ante} perception", which is the
individual presumption of the satisfaction that a good can provide;
and (2) an ``\emph{ex-post} evaluation", which is the satisfaction
that a good existing on the market actually provides. In this
respect, the level of satisfaction associated to a given node of the
graph represents what we call its ``ex-ante utility'', which is
just an estimate of its ``effective utility'' (measurable only after
the consumer physically explore the node). We model this fuzzy
preference structure by a \emph{total preorder} on the set of
nodes.\footnote{Recall that a \emph{total preorder} on a set $X$ is
a binary relation $\succsim$ on $X$, which is \emph{reflexive} ($x
\succsim x$ for each $x \in X$), \emph{transitive} ($x \succsim y$
and $y \succsim z$ imply $x \succsim z$ for each $x,y,z \in X$), and
\emph{complete} ($x \succsim y$ or $y \succsim x$ for each $x,y \in
X$ such that $x \neq y$). In this case, the \emph{indifference
relation} $\sim$ associated to $\succsim$, defined by $x \sim y$ if
$x \succsim y$ and $y \succsim x$, is an equivalence relation on
$X$.} This is a classical hypothesis in individual preference
theory. Since the number of nodes is finite, the advantage of this
framework is that it admits a \emph{utility representation} with
``thick'' indifference classes.\footnote{ A binary relation
$\succsim$ on $X$ is \emph{representable} if there exists an
\emph{order-embedding} from $X$ into the set $\mathbb{R}$ of real
numbers, i.e., a map $u \colon X \to \mathbb{R}$ such that for each
$x,y \in X$, we have $x \succsim y$ if and only if $u(x) \geq u(y)$.
In this case, the function $u$ is called a \emph{utility
representation} of $\succsim$. It is well-known that a total
preorder on a countable set (hence, in particular, on a finite set)
is always representable: see, e.g., \cite{bridges2013representations,
aleskerov2007utility} for a general discussion about utility representations
and some technical results.}

Figure~2 provides a graphical description of such a consumer's
preference structure. The $\mathbb{R}^2$ space endowed with the
satisfaction metric is partitioned into $N_\Lev$ equivalence
classes, called \emph{indifference levels} and represented by
concentric annuli centered at the target. All the nodes in an
annulus are \emph{ex-ante} equally preferred by the consumer, i.e.,
they display the same \emph{presumed} utility:\footnote{Formally, if
$\succsim$ is the total preorder on  $X$ representing the preference
structure of a consumer, $\sim$ is the indifference associated to
$\succsim$, and $u \colon X \to \mathbb{R}$ is an utility
representation of $\succsim$, then we have $x \sim y$ if and only if
$u(x) = u(y)$. So each annulus can be seen as a fuzzy
multidimensional representation of indifference curves.} the further
from the target an annulus is, the lower the presumed utility
associated to the corresponding indifference level is, and vice
versa. Clearly, a low value of $N_\Lev$ is typical of a consumer
with a rather low ex-ante discrimination power, whereas a large
value of $N_\Lev$ indicates a consumer with a high ability in
discriminating among similar goods.
\end{description}

As a conclusive remark, note that we do not assume \emph{a priori}
the existence of a relationship between knowledge and discrimination
ability. Therefore, we can identify several ``categories of
buyers'', which display various combinations of these two features.
In the next subsection we illustrate how consumers interact with the
market structure, and how this interaction -- along with some
additional individual features -- determines different categories of
consumers.


\subsection{Market-Consumer Interaction}
\label{SUBSECT:gravitational field}

Each central node (hub) of the graph can be considered as a ``pole
of attraction'' for the consumer. In the case that hubs represent
brands, the strength of such attraction depends on many factors,
such as marketing suggestions, variety of offered products,
advertising and promotional campaigns, communications and
brand-management policies, etc. To simplify our analysis, in this
paper we assume that the attractive power of a hub is only a
function of the amount of products having that brand. In fact -- in
analogy with the gravitational field of a body in physics -- we
introduce the \emph{mass} $M$ of a hub, which is by definition equal
to the number of leaves of the corresponding cluster. Note that the
attraction effect generated by each hub becomes effective only if
the consumer is aware of the presence of the corresponding cluster:
said differently, ``knowledge determines attraction''.

In Figure 3 we graphically represent the \emph{attraction field} induced
by the three active clusters of Figure 2 (the four inactive clusters
have no effect whatsoever on the consumer, since she  cannot even
perceive their existence). For the sake of visualization, only the
central nodes are reported in the figure. The integer number next to
each hub is its mass $M$, which provides a discrete measure of the
strength of directional attraction. The arrows of the attraction field
may possibly push the consumer to choose in a certain direction.
However, not every consumer will be affected in the same way by the 
field: in fact, as we shall argue later on, the more a
consumer is well aware of her personal tastes/goals, the less she  is
influenced by market suggestions. In order to take into account this
kind of behavior, next we introduce the last feature that
characterizes a consumer in our model.

\begin{description}
  \item[\bf Awareness.] This is the capability of the consumer to
  discern the features of a product without being subject to the
  influence of the market: the higher her awareness,
  the less likely that she  makes her choice
  according to the market's attraction field.
  In fact, a highly aware consumer will always take advantage of both her
  knowledge and her ex-ante discriminating power to effectively
  explore the market while moving towards the target. In our model
  we assume that awareness is a parameter $\Aw$ that varies in the closed interval $[0,1]$.
\end{description}

\begin{figure}
\centering
\includegraphics[width=0.5\textwidth]{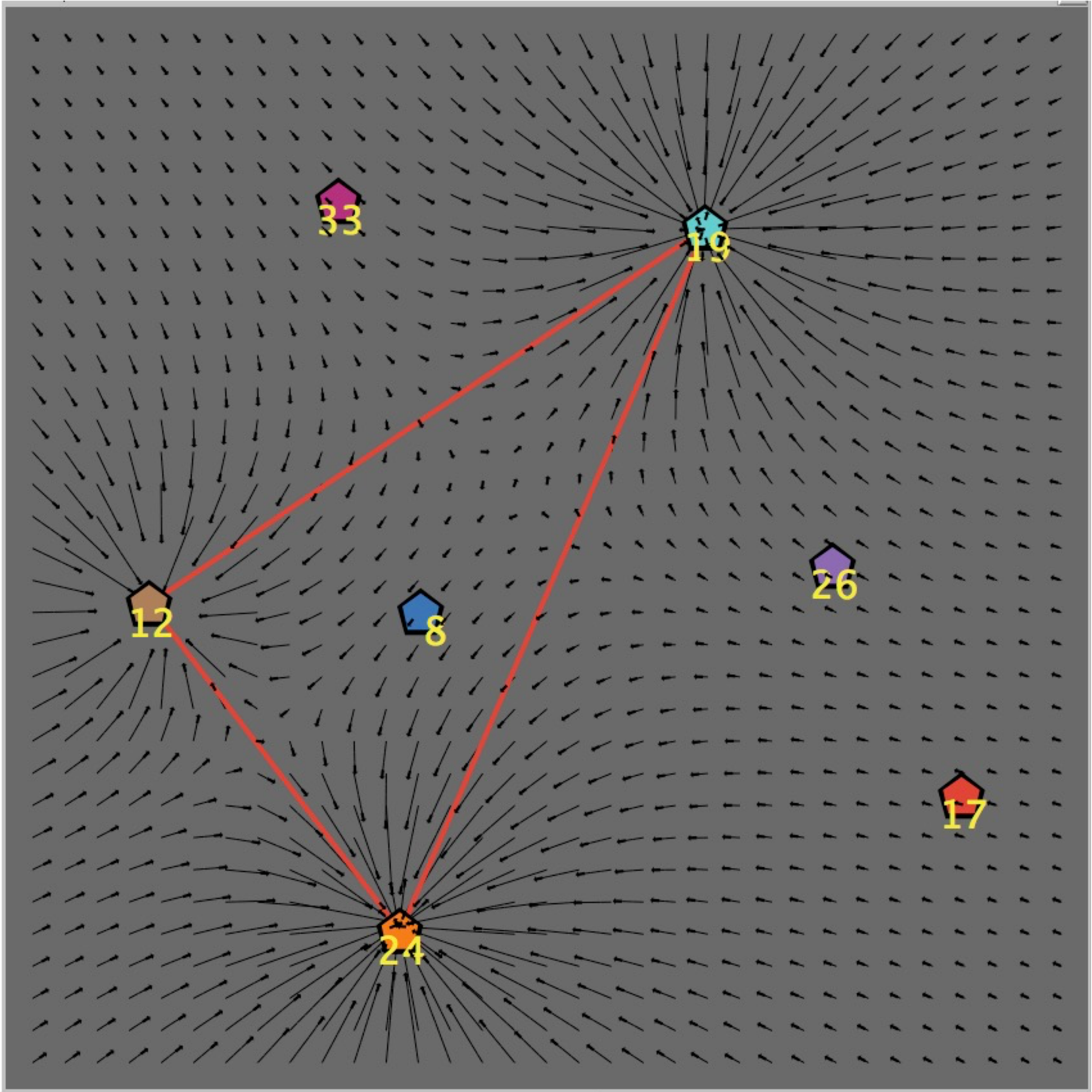}
\caption{\footnotesize \emph{Market-consumer interaction: attraction
field induced by three active hubs.} The label next to each hub
represent its mass, which is equal to the number of products
(leaves) of the corresponding cluster. All intermediate and terminal
nodes are not represented.}
\end{figure}

Knowledge ($\Kn$), discriminating ability ($N_\Lev$) and awareness
($\Aw$) are the three parameters that allow us to distinguish among
several categories of consumers, characterized by suitable
combinations of these features. In Table $1$ we list a taxonomy of
cases, in which the extreme values of the parameters better
emphasize the differences among different types of buyers. In
particular, consumers from type \#1 to type \#5 (with $\Kn=0.5$)
display an imperfect knowledge of the market, while consumers from
type \#6 to type \#10 (with $\Kn=1$) know the market perfectly.
Within each of these two general categories, we have the extreme
cases of consumers with either minimum ($\Aw=0$) or maximum
($\Aw=1$) awareness, as well as a special type of consumer who -- as
we shall describe later on -- walks along her informative path
completely {\it at random}. Finally, for non-random cases, we
consider consumers with low and high discriminating ability
(respectively, $N_\Lev = 4$ and $N_\Lev = 20$).

These ten combinations of the main individual parameters describe a
fictitious community made by several distinct (non-interacting)
consumers, who aim at maximizing their own utility according to
their subjective awareness, discrimination capacity and
knowledge. As we explain in detail in the next section, each type of
consumer will navigate the graph in search of her (existing or
non-existing) target, following step by step (i.e., node by node) a
personal informative path starting at the source (where she  is
located at time $t=0$).

\bigskip

\begin{table}[h]
\begin{center}
\small
\begin{tabular}{|c||c|c|c|}
\hline
Consumer &  $\Kn$  &   $\Aw$   & $N_\Lev$ \\
\hline \hline
\#1        &  $0.5$ &   $0$   &   4   \\
\#2        &  $0.5$ &   $0$   &   20  \\
\#3        &  $0.5$ &   $1$   &   4   \\
\#4        &  $0.5$ &   $1$   &   20  \\
\#5        &  $0.5$ &  \emph{random} &   $-$ \\
\hline
\#6        &  $1$  &   $0$   &   4   \\
\#7        &  $1$  &   $0$   &   20  \\
\#8        &  $1$  &   $1$   &   4   \\
\#9        &  $1$  &   $1$   &   20  \\
\#10       &  $1$  &  \emph{random} &   $-$ \\
\hline
\end{tabular}
\end{center}
\caption{\footnotesize \emph{Ten types of consumers characterized by
different values of individual parameters.} Only rather extreme
values of knowledge ($\Kn$), awareness ($\Aw$) and discrimination
ability ($N_\Lev$) are used. The special type of random consumer
also appears.}
\end{table}

Next, we anticipate the possible outcomes of the informative journey
of a generic consumer. Let $t_{\mathrm{end}}$ be the total number of
explorative steps done by the consumer, who starts her journey at
time $t_0$. Further, let $P$ be the product (present in the market)
delivering the maximum satisfaction among the ones visited by her,
and $t^*$ the step at which $P$ is reached. (Thus, time is discrete,
and $t^*$ is an integer between $t_0$ and $t_{\mathrm{end}}$.) If
$\Sat(P,t^*)$ denotes such a maximum satisfaction, then the
individual journey may give rise to four possible outcomes for the
consumer, which depend on whether the target product exists (cases
1a and 1b) or not (cases 2a and 2b).
\begin{itemize}
  \item[(1a)] \emph{The target coincides with a real node of the graph,
  and is actually reached.}
  In this case, the consumer ends her search at the target $P^*$ (i.e., $P = P^*$ and $t^* = t_{\mathrm{end}}$),
  and, following Equation~(\ref{satisfaction}), her satisfaction is maximum (i.e., $\Sat(P^*,t^*)=\Sat_{\max}=1$).
  \item[(1b)] \emph{The target coincides with a real node of the graph,
  but is not reached.} In our model there are no constraints on the time that each buyer can spend
  while searching for her target.
  However -- as we better explain in the next section -- the consumer might eventually get trapped
  in an ``informative cul-de-sac'', in which case she  will buy the product corresponding to the
  maximum satisfaction reached until that moment. Specifically, if this maximum satisfaction is given by a certain product $P$
  reached at time $t^*$, then the consumer's satisfaction $\Sat(P,t^*)$ is a number in the open interval $(0,1)$.
  \item[(2a)] \emph{The target does not coincide with a real node of the graph,
  but the consumer reaches a node at minimum distance from it.}
  Whenever the ideal product does not exist in the market, choosing the ``closest'' product to it (at, say, time $t^*$)
  is actually the best that the consumer can do. As a consequence, her satisfaction
  ``should'' theoretically be maximum, even if she  will inevitable end
  her journey in a cul-de-sac. However, unless the consumer has a
  precise perception that the ideal product is not present in the market, at the end of her journey she  will
  still believe that her original goal has not been fully
  accomplished. Therefore, as in case (1b), we have again
  $0<\Sat(P,t^*)<1$.
  \item[(2b)] \emph{The target does not coincide with a real node of the graph,
  and the consumer does not even reach a node that has the minimum distance from it.}
  In this case, there are no doubts that the consumer's satisfaction
  is not maximum (in fact, lower than in the previous case).
  As in cases (1b) and (2a), her final satisfaction
  corresponds to the maximum one reached (at, say, time $t^*$) during her exploration of the
  market, before being (inevitably) trapped in a cul-de-sac. Again, we have $0<\Sat(P,t^*)<1$.
\end{itemize}

Of course, the outcome effectively realized is also determined by
several factors deriving from the heterogeneity of individuals, such
as the effects of advertising, a lack of knowledge, a very
inaccurate ex-ante preference structure, etc. Therefore, we expect
that distinct types of consumers -- endowed with different levels of
knowledge, awareness and discriminating power -- will behave quite
differently. For instance, it is apparent that only consumers with
$\Kn=1$ (from type $\#6$ to type $\#10$ in Table 1) will have the
possibility to reach the target, provided that the latter is a real
product on the market (outcome (1a)). On the other hand, a lack of
connections between central nodes due to medium values of $\Kn$
(consumers from type $\#1$ to type $\#5$ in Table 1) will likely
hinder the achievement of the target, even in cases in which the
latter corresponds to a product existing in the market (outcome
(1b)).


\subsection{Dynamics of the model} \label{SUBSECT:dynamics_of_model}

We now describe how a consumer effectively moves from a given source
toward her target. This will enable us to evaluate differences among
types of consumers in a statistically significant way.

To start, note that the fact that our consumer ``has in mind'' her
target does not imply any knowledge about its exact location with
respect to the nodes of graph. As a matter of fact, in the (rather
unlikely) circumstance that the buyer possesses this perfect
knowledge, she  would immediately select an existing product that is
as close as possible to the target, and there would be no
explorative process. Instead, our model aims at describing the whole
``consumption experience'' from a dynamical perspective.

Actually, we assume that, as soon as the need of a certain good is
perceived,\footnote{Here we do not explicitly address the question about 
the ``repeated consumption" experience.} the consumer has a (more or less) vague idea of what \emph{would}
satisfy her best. Thus, she  starts her informative journey in order
to find it, looking at the market structure according to her
knowledge $\Kn$, and inspecting existing goods in a way that is
influenced by her awareness $\Aw$ and her discrimination ability
$N_\Lev$. Specifically, the consumer starts at time $t=0$ from a
randomly chosen source node belonging to one of her active clusters,
and moves at time $t=1$ to another node of the graph, following
either a yellow link (connecting nodes of the same cluster) or a red
link (going from the hub of a cluster to another hub). At any time
step $t$ the consumer is on a certain node of the graph, and at
time $t+1$ she  moves to another node of the graph, following either
a yellow link or a red link. To describe the algorithm that guides
the consumer's selection at each step of her journey on the graph,
let us assume that at time $t$ the consumer is on a given node $j$
with degree $k_j$. Now the question is: \emph{how does the consumer select the neighbor node where to go at time $t+1$?}.
In what follows, we give two distinct answers to this question,
distinguishing between the two main categories of consumers: (1)
non-random; (2) random.
\begin{figure}
\centering {\includegraphics[width=0.4\textwidth]{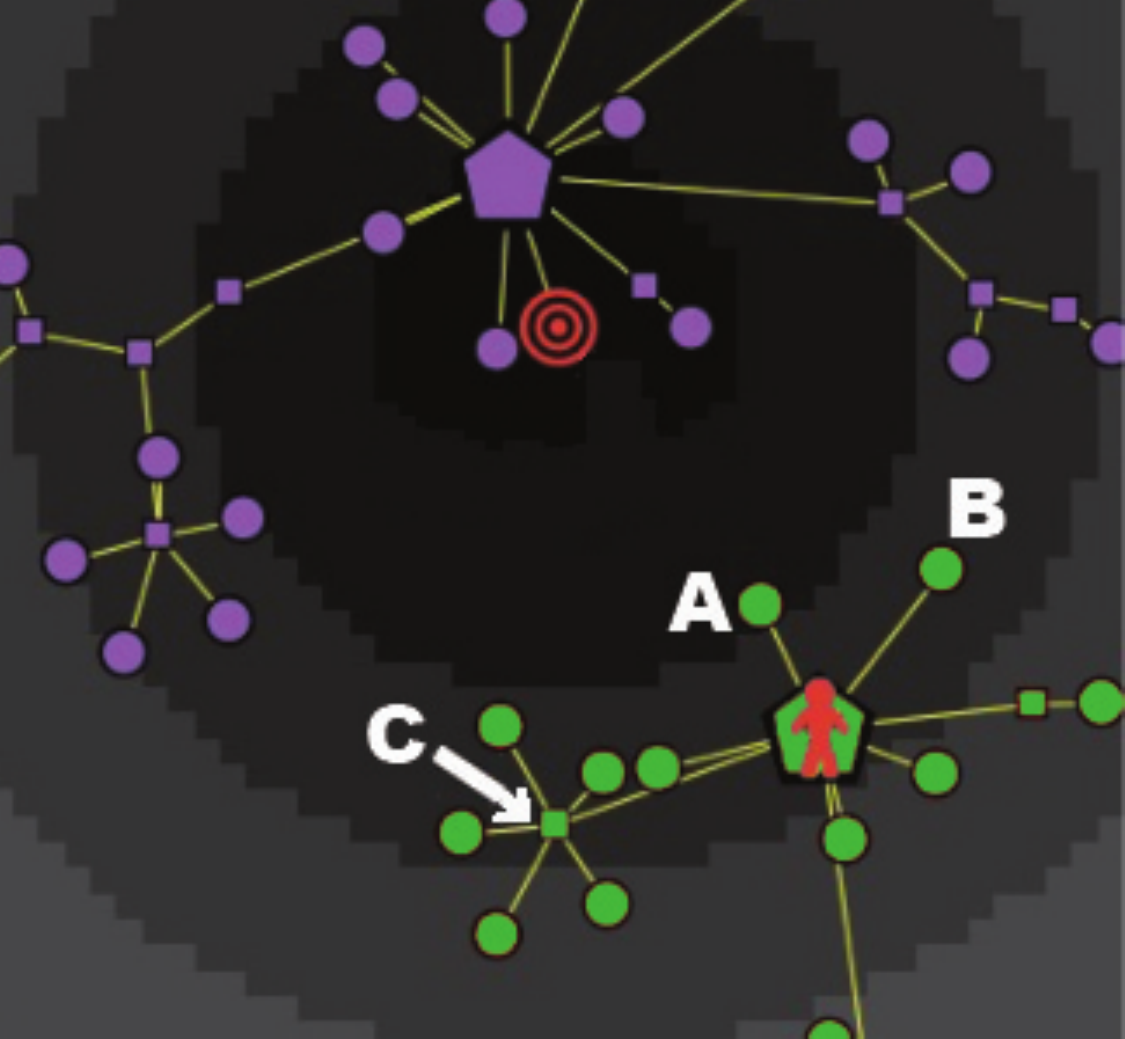}}
\caption{\footnotesize \emph{Dynamical rules determining the buyer's
next move: the case of high awareness.} The consumer is well aware
of the target, and tries to approach it by exploiting her high
discrimination ability.}
\end{figure}
\\
\\
(1) For \emph{non-random} consumers, the answer to the above
question strictly depends on the values of the parameters $\Aw$ and
$N_\Lev$. In fact, a consumer who is quite aware of her active
segment of the market (i.e., of all the products belonging to her
active clusters) will theoretically be guided only by her attempt to
approach the target; on the other hand, the moves of a scarcely
aware consumer will be largely influenced by the attraction field
induced by her active clusters, almost independently of the position
of the target. We translate these two opposite tendencies into,
respectively, the ``dynamical rules'' (a) and (b) described below.

\begin{itemize}
\item[(a)] \emph{With probability $p=\Aw$, the consumer chooses
one the first neighbors with highest degree among those with highest
utility (according to her preference structure).} More precisely,
first the consumer ranks all ``reachable'' indifference levels by
the total preorder that models her preference structure, that is, he
determines the indifference level $L_i$ containing a first neighbor
that is the closest to the indifference level $L_0$ of the target.
If there is only one first neighbor in $L_i$, then she  chooses it.
Otherwise, she  chooses randomly one the first neighbors in $L_i$
among those having maximum degree.\footnote{In the majority of
cases, there will be a unique first neighbor with maximum degree in
the best reachable indifference class.} The rationale guiding this
selection process is very natural: the consumer selects according to
a criterion of highest possible utility, and, in cases of
\emph{ex-equo}, she  makes her choice according to the highest degree,
since this selection increases her freedom of movement at the very
next step of the informative process.
\\
In Figure 4 we represent the case of a highly aware consumer: at
time $t$, she  is on the hub of a cluster, located in an indifference
class that is two levels below the best. According to our algorithm,
at time $t + 1$ the consumer will choose (with probability $p=\Aw$)
one of the first neighbors that belongs to the best reachable
indifference class. In our case, there is only one node of this
kind, namely $A$. Note that if node $A$ were not to be present in
the graph, then the consumer would have chosen node $C$, since
this is the node with highest degree among the first neighbors in
the best indifference level. For example, in Figure 4, the two nodes
$C$ and $B$ belong to the same indifference class, but $C$ (with
degree $6$) is better than $B$ (with degree $1$).

\begin{figure}
\centering {\includegraphics[width=0.445\textwidth]{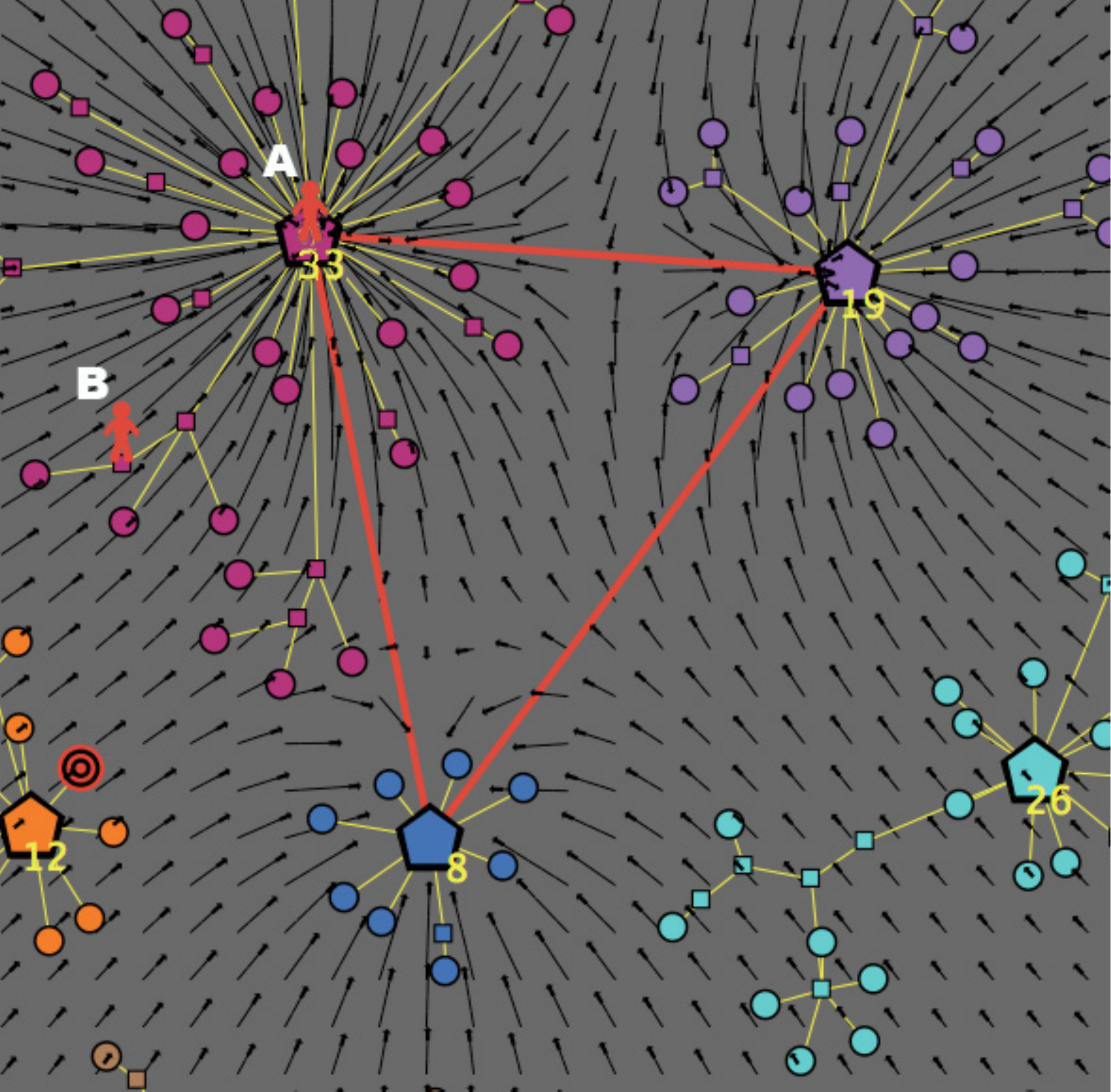}}
\caption{\footnotesize \emph{Dynamical rules determining the buyer's
next move: the case of low awareness.} The consumer is highly
influenced by the attraction field induced by her active clusters,
without consideration for the target.}
\end{figure}

\item[(b)] \emph{With probability $p=1-\Aw$, the consumer select a first neighbor on the
basis of the attraction field, without any consideration for the
target.} In the example shown in Figure 5, a scarcely aware consumer
is influenced by the attraction generated by the three active
clusters (having masses equal to, respectively, $33$, $19$ and $8$,
as shown by the corresponding labels), whereas the target is located
on another (unknown) cluster. In this case, the selection depends on
the position $j$ of the consumer at time $t$, according to the
following two subcases:

$\bullet$ if at time $t$ the consumer is on a central node (position
$A$ in the figure), then at time $t+1$ she  will move to one of her
$k_j$ first neighbors \emph{with a probability proportional to the
mass of each node}\footnote{For nodes that are not hubs, the mass
simply corresponds to their degree.}; obviously, the central nodes
of active clusters with a large number of leaves have a very high
probability to attract the buyer (in our example, the central node
with mass $19$ will most likely be the buyer's next choice);

$\bullet$ if at time $t$ the consumer is on an intermediate or a
terminal node (position $B$ in the figure), then at time $t+1$ he
will move \emph{at random} to one of her $k_j$ first neighbors; in
this way, the consumer has the opportunity to explore all the
informative trees in her active clusters, without being forced to
come back to the central nodes (which maximally attract her).
\end{itemize}

\noindent (2) On the other hand, a consumer may decide to walk along her
informative path \emph{completely at random}, that is, following the
(red or yellow) links of her active clusters according to no
predetermined rule. In this case, at each time $t$ the buyer will
choose randomly one of her $k_j$ first neighbors, and move toward it
at time $t+1$. One of the most remarkable results of this paper is
that the random strategy is far from being a losing one. In fact, it
turns out that a random walk over the (active clusters of the)
graph will result more effective, in terms of utility, than the
analogous one influenced only by the attraction field (see Figure 5,
in the extreme case $\Aw=0$). This unexpected result is extensively
discussed later in the paper.
\begin{figure}[t]
\centering
\includegraphics[width=0.45\textwidth]{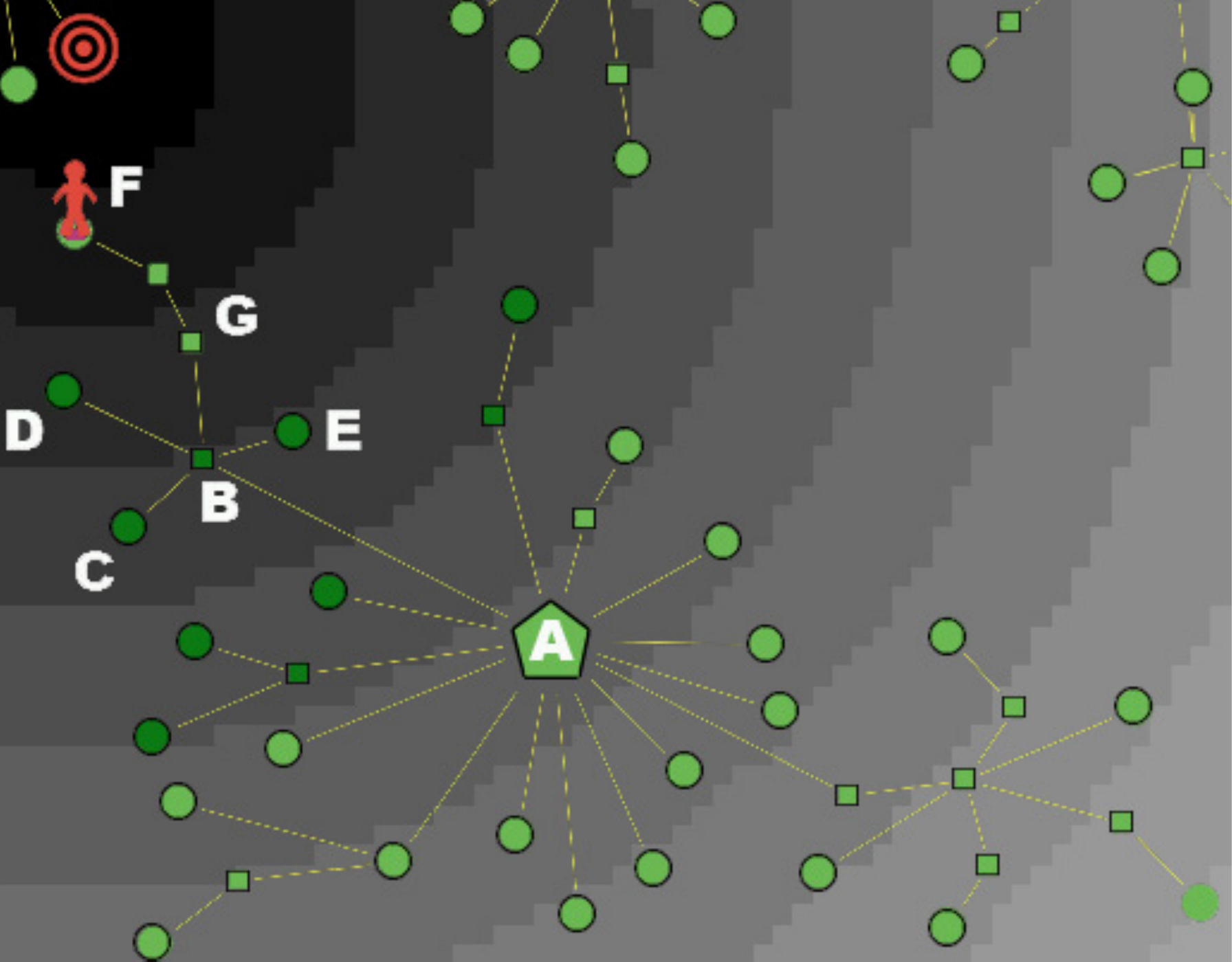}
\caption{\footnotesize \emph{Example of a cul-de-sac.} In an attempt
to reach her target, the buyer is exploring a branch of her active
cluster, but remains trapped in an informative cul-de-sac, and will
end her journey on node $G$.}
\end{figure}

\bigskip

Finally, we describe the termination of the algorithm, that is, how
a given type of consumer may end her informative journey. In this
respect, we need to point out an additional feature of the selection
process, which affects the dynamical rules described above. In fact,
the algorithm puts a constraint on the number of times that a buyer
can visit a given node of her active subgraph: this upper bound is
reasonably set as equal to the degree of the node itself. We impose
this limitation in accordance to the (realistic) assumption that a
consumer should not travel along the same path more than once.
Therefore, after $k$ visits, a node with degree $k$ will be
``switched-off'', and it will be no longer possible to cross it. 

In Figure 6 we show a buyer (with awareness $\Aw \sim 0.5$) who,
starting from the central node $A$, passes through the intermediate
node $B$ in order to visit the terminal nodes $C$, $D$ and $E$. All
visited nodes at a certain stage are colored in dark green, which
stands for ``switched-off''. Now imagine that, before going towards
node $F$, the consumer comes back one more time to node $A$ (for
example, because she  is moving at random). When she  finally walks
towards $F$, node $B$ (which has degree $k=5$) will be switched-off
too, and the buyer's fate is to remain trapped in a cul-de-sac: in
fact, she  will end her journey on node $G$. As a consequence of this
feature of the algorithm, the consumer can terminate her journey (at
time $t=t_{\mathrm{end}}$) in two different ways, described below.
\begin{itemize}
\item[(a)] \emph{The consumer remains trapped in an informative cul-de-sac.}
(This situation may happen regardless of the fact that the target
coincides with an existing node of the graph or not.) In this
case, the consumer buys, among the products visited until the time
$t_{\mathrm{end}}$, the product $P$ corresponding to the maximum
relative satisfaction, and so $\Sat(P,t^*)<1$. The chosen product
$P$ is located at the relative minimum distance $d(P,P^*)$ from the
target $P^*$ (see Equation 1).\footnote{Note that, since we are in the
metric space $\mathbb{R}^2$ with the Euclidean distance, the visited
node with minimum distance from the target is essentially unique
from a probabilistic point of view.}

\item[(b)] \emph{The consumer reaches her target and buys it.} This
can happen only if the target coincides with an existing node that
belongs to one of the active clusters. In this case, the algorithm
terminates at the target $P^*$, and the satisfaction of the consumer
is given by $\Sat(P^*,t_{\mathrm{end}})=1$.

\end{itemize}

We end this section by defining two measures of the consumer's
performance in the whole process.
The first one is the \emph{final utility} $U$, defined as the
maximum value of satisfaction obtained by the consumer at time
$t^*\in [0,t_{\mathrm{end}}]$ during her informative journey, that
is,
\begin{equation}
U := \Sat(P,t^*)
\end{equation}
The second one is the \emph{total efficiency} $H$ of the consumer's
experience, defined as
\begin{equation}
H := \frac {t^*}{t_{\mathrm{end}}}.
\end{equation}

These definitions readily yield that the equality $(U,H) = (1,1)$
only holds in one case, namely, whenever the buyer is able to reach
his target (at time $t^*=t_{\mathrm{end}}$). In all other cases, we
will have $0 \leq U < 1$ and $0 \leq H < 1$, with values depending
on the individual parameters $\Kn$, $N_\Lev$ and $\Aw$.

We aim at characterizing the ten types of consumers described in
Table 1 by their respective values of $U$ and $H$. Note that each
simulation run of the algorithm corresponds to a single informative
journey with fixed values of $\Kn$, $N_\Lev$ and $\Aw$. Further, the
performance of each type of consumer is obviously influenced by the
random initial positions of both the buyer and the target, as well
as by the random selection of the active clusters. Therefore, in
order to obtain statistically significant result, we need to
calculate $U$ and $H$ for each category of buyers over many
different simulation runs (events), starting from different initial
conditions. The next section is devoted to a detailed description of
the implementation of this procedure.


\section{Simulation Results} \label{SECT:simulations}

Here we present and discuss the obtained results. 
The market structure considered in the simulations is the one represented in Figure 2, with seven clusters embedded in a satisfaction space.

For the first five types of consumers described in Table 1 (with
limited knowledge $\Kn=0.5$), there are three active clusters (out
of seven). For each type of consumer, we consider a set of $50$
simulations, called {\it multievents}, with different random choices
of the three active clusters.\footnote{Note that $50$ simulations
suffice since we have $\binom{7}{3} = 35$ possible choices of three
clusters out of seven.} Further, for each multievent we perform a
set of $500$ simulations, with different initial positions of both
buyer and target, for a total of $N=25000$ events. Finally, over the
whole set of $N$ events, we compute the distributions $P(U)$ and
$P(H)$ of final utility $U$ and total efficiency $H$, respectively.
The mean value of the final utility distribution is called {\it
Social Welfare}, and is denoted by $W$; its standard deviation
$\sigma(W)$ is reported, too. We also calculate the distribution
$P(t_{\mathrm{end}})$ of stopping times, and report its mean value $T=
\langle t_{\mathrm{end}} \rangle$ and the corresponding standard
deviation $\sigma(T)$. For statistical significance, a similar set
of $N=25000$ simulations is also done for consumers with complete
knowledge, corresponding to types from $\#6$ to $\#10$ in Table
1.\footnote{Indeed, since $\Kn = 1$ implies that the seven clusters
of the market are all active and reciprocally connected, there would
be no need of averaging over the $50$ randomly chosen different
subsets of the active clusters. We repeat the experiment only for
the sake of completeness.}

In what follows, we discuss the results of the simulations,
separating the analysis of the following two cases: (1) \emph{target
on product}, that is, the target coincides with a product existing
on the market; and (2) \emph{target off product}, that is, the
target is only ideal, since there is no product on the market that
perfectly corresponds to the consumer's goal.


\begin{figure}
\centering \subfloat [][~~~Distributions for consumers 1 to 5
($\Kn=0.5$)] {\includegraphics[width=0.485\textwidth]{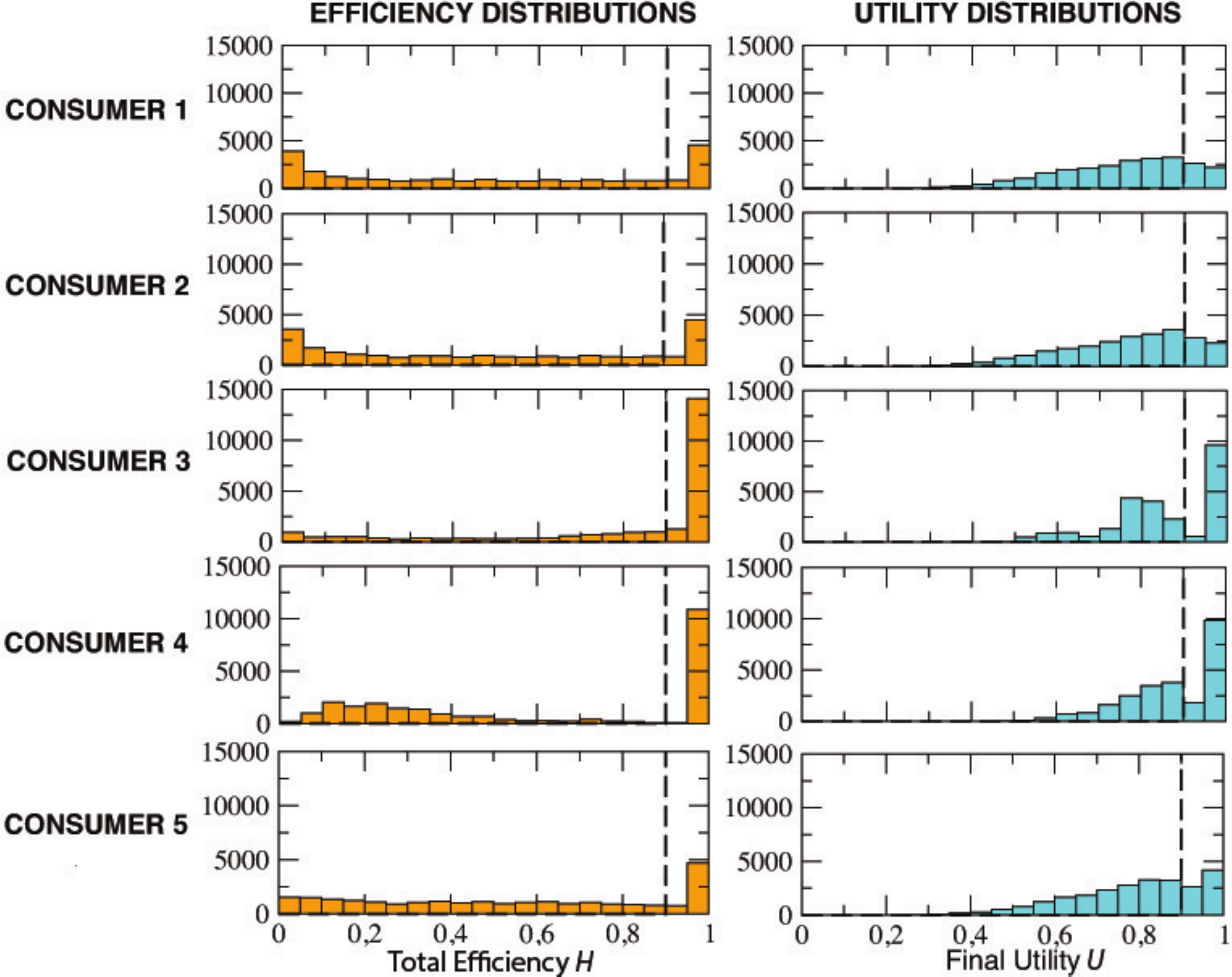}}
\quad \subfloat [][~~~Distributions for consumers 6 to 10
($\Kn=1.0$)] {\includegraphics[width=0.485\textwidth]{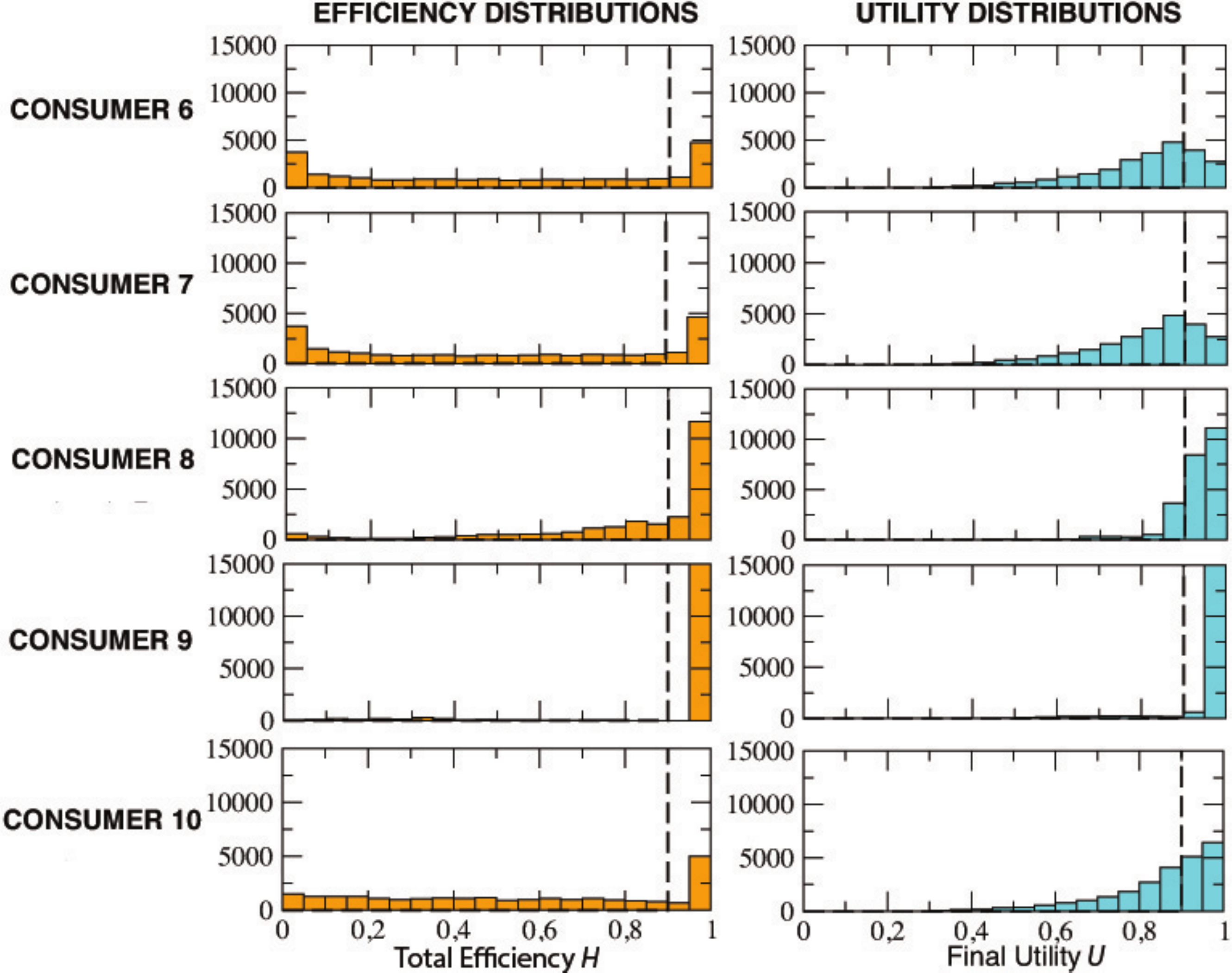}}
\caption{\footnotesize \emph{Market with ``target on product''.}
Distributions of efficiency and final utility calculated over
$N=25000$ simulation events for the 10 types of consumers shown in
Table 1. Dashed lines indicate levels of efficiency and utility
higher than 0.9.}
\end{figure}

\subsection{Target on product}

Assume that the target coincides with one of the terminal nodes of
the graph. In this situation, consumers can in principle reach the
target without remaining trapped in some informative cul-de-sac;
however, this result strictly depends on their knowledge, awareness
and discriminating ability. In Figure 7 we plot the distributions of
both total efficiency and final utility for the 10 types of
consumers. In particular, in panel (a) we consider consumers with
limited knowledge ($\Kn=0.5$), whereas in panel (b) we describe
consumers with complete knowledge ($\Kn=1$).

It is expected that group (b) performs better than group (a), with
respect to both efficiency and utility. It is also far from
surprising that perfectly aware buyers ($\Aw=1$) are in a better
position than totally unaware ones ($\Aw=0$). In particular, only
type $\#9$ -- who is endowed with a perfect knowledge and awareness,
and has a high discriminating ability $N_\Lev=20$ -- is able to
reach her target in almost all the $25000$ events, thus totaling the
highest scores (equal to $1$) for both $H$ and $U$.\footnote{Note
that, however, the $y$-axis of her plots is truncated at $15000$.}

On the contrary, it is rather unexpected that random consumer types
($\#5$ and $\#10)$ are in a better position than the corresponding
unaware ones ($\#1$, $\#2$, $\#6$ and $\#7$), since their
distributions are more shifted to the right. However, this seemingly
strange effect can be rationally explained after a moment's thought.
Indeed, a consumer who moves around the graph only according to
the attractiveness of various brands, with no awareness of the
market and no discriminating ability whatsoever, is likely to be
trapped into a ``\emph{branding loop}'': she  blindly wonders around
the graph, being pulled and pushed in all directions by the
strength of each brand, and ends up loosing sight of her original
goal. Under the same hypothesis, moving completely at random may
result into a winning strategy, since independence from the
attraction field will likely allow the consumer to explore quite a
few nodes before stopping her informative journey.

To better appreciate the differences in performances among the
various types of consumers, in Table 2 we list the following values:
the percentage of events with final utility $U>0.9$ and the
percentage of events with total efficiency $H>0.9$ (see dashed lines
in all the plots of Figure 7), the social welfare $W$ with its
standard deviation $\sigma(W)$, and the average stopping time $T$ with
its standard deviation $\sigma(T)$.

\begin{table}[ht]
\begin{center}
\small
\begin{tabular}{|c|||c|c|c||c|c||c|c|c|c|}
\hline Consumer &  $\Kn$  &   $\Aw$   & $N_\Lev$  &  $H(\%)>0.9$ &
$U(\%)>0.9$ & $W$ &
$\sigma(W)$ & $T$ & $\sigma(T)$\\
\hline \hline
\#1        &  $0.5$ &   $0$   &   4   &  $21$ &   $19$   &   $0.76$    &  $0.16$ &   $46$ &   $24$   \\
\#2        &  $0.5$ &   $0$   &   20 &  $21$ &   $20$   &   $0.76$  &  $0.15$ &   $46$ &   $24$  \\
\hline
\#3        &  $0.5$ &   $1$   &   4   &  $61$ &   $41$   &   $0.86$    &  $0.13$ &   $25$ &   $18$    \\
\#4        &  $0.5$ &   $1$   &   20 &  $44$ &   $47$   &   $0.88$   &  $0.11$ &   $21$ &   $18$    \\
\hline
\#5        &  $0.5$ & \emph{random} &   $-$ &  $22$ &   $27$   &   $0.79$  &  $0.15$ &   $43$ &   $24$   \\
\hline
\hline
\#6        &  $1$  &   $0$   &   4    &  $23$  &   $27$   &   $0.80$   &  $0.14$ &   $75$ &   $42$     \\
\#7        &  $1$  &   $0$   &   20  &  $23$  &   $27$   &   $0.80$  &  $0.14$ &   $76$  &   $42$   \\
\hline
\#8        &  $1$  &   $1$   &   4   &  $56$  &   $78$   &   $0.93$  &  $0.07$ &   $14$  &   $14$   \\
\#9        &  $1$  &   $1$   &   20  &  $93$  &   $94$   &   $0.97$  &  $0.07$ &   $7$  &   $5$   \\
\hline
\#10      &  $1$  & \emph{random} &   $-$&  $23$  &  $46$   &   $0.85$  &  $0.14$ &   $60$  &   $39$    \\
\hline
\end{tabular}
\end{center}
\caption{\footnotesize \emph{Market with ``target on product''.}
Global quantities calculated over the $N=25000$ simulation events
for the 10 types of consumers described in Table 1. We report, in
order, the following quantities: the percentage of events with final
utility $U>0.9$, the percentage of events with total efficiency
$H>0.9$, the social welfare $W$ with its standard deviation
$\sigma(W)$, and the average stopping time $T$ with the corresponding
standard deviation $\sigma(T)$.}
\end{table}

As expected, the outcomes of both random consumers and those with no
awareness do not depend on the number $N_\Lev$ of indifference
levels, apart from insignificant statistical fluctuations. Table 2
also confirms a better performance of the group with total knowledge
($\Kn=1$) over that with partial knowledge ($\Kn=0.5$). Within these
groups, it is apparent the superiority of totally aware consumers
($\Aw=1$) and, in particular, of consumer $\#9$, not only concerning
$H$, $U$ and $W$, but also regarding the average simulation time
$T$:
indeed, consumer $\#9$ reaches the target almost always ($W=0.97$)
and, in average, very quickly ($T=7$), thus confirming the
effectiveness of our algorithm in conditions of perfect knowledge.
Finally, data confirms the better score of random consumers ($\#5$
and $\#10$) with respect to the completely unaware ones (e.g., $\#2$
and $\#7$), in particular for what concerns final utility.

Figure 8 describes the behavior of efficiency $H$ and utility $U$ as
a function of the consumer's awareness $\Aw$. Specifically, the
outcomes of $H(\%)>0.9$ and $U(\%)>0.9$ are reported for several
values of awareness, fixed $\Kn=1$ and $N_\Lev=20$; for the sake of
comparison, results for random consumers are also listed. Note that
scores increase along with $\Aw$ and rapidly saturate, reaching
already their maximum value at $\Aw=0.5$. The interpretation of
these results is rather natural. On one hand, there is no need to be
perfectly informed consumers in order to get the maximum
satisfaction from a purchase: in fact, a medium amount of
information suffices. On the other hand, for scarcely informed
consumers, a small quantity of information about the market is
enough to overcome the performance of a random buyer. However,
should one also take into account the cost needed to gather
information, it is likely that a random search strategy would gain a
better position in the total ranking.

\begin{figure}
\centering \subfloat [][Percentage of events with total efficiency
$H(\%)\!>\!0.9$]
{\includegraphics[width=0.485\textwidth]{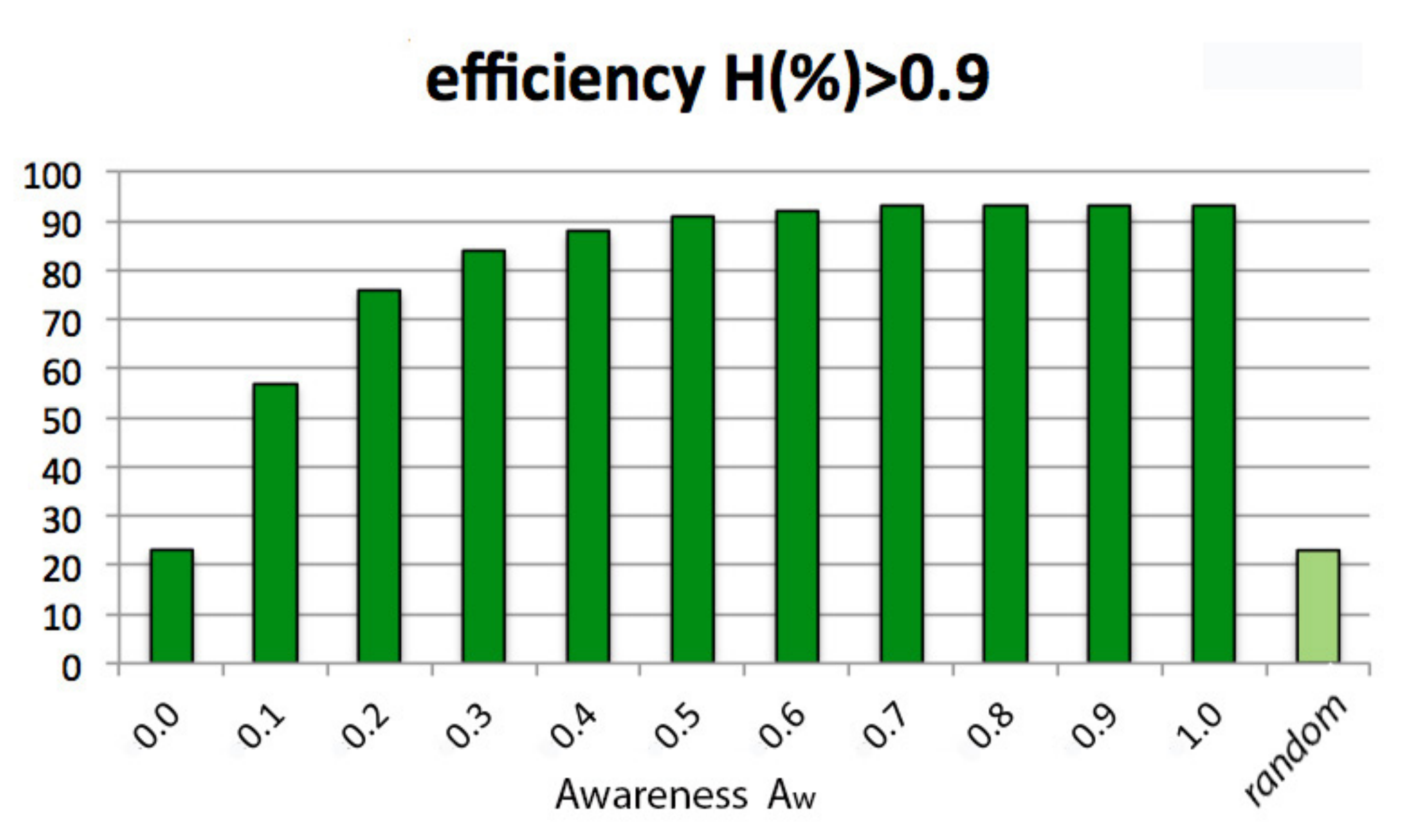}} \quad
\subfloat [][Percentage of events with final utility
$U(\%)\!>\!0.9$]
{\includegraphics[width=0.485\textwidth]{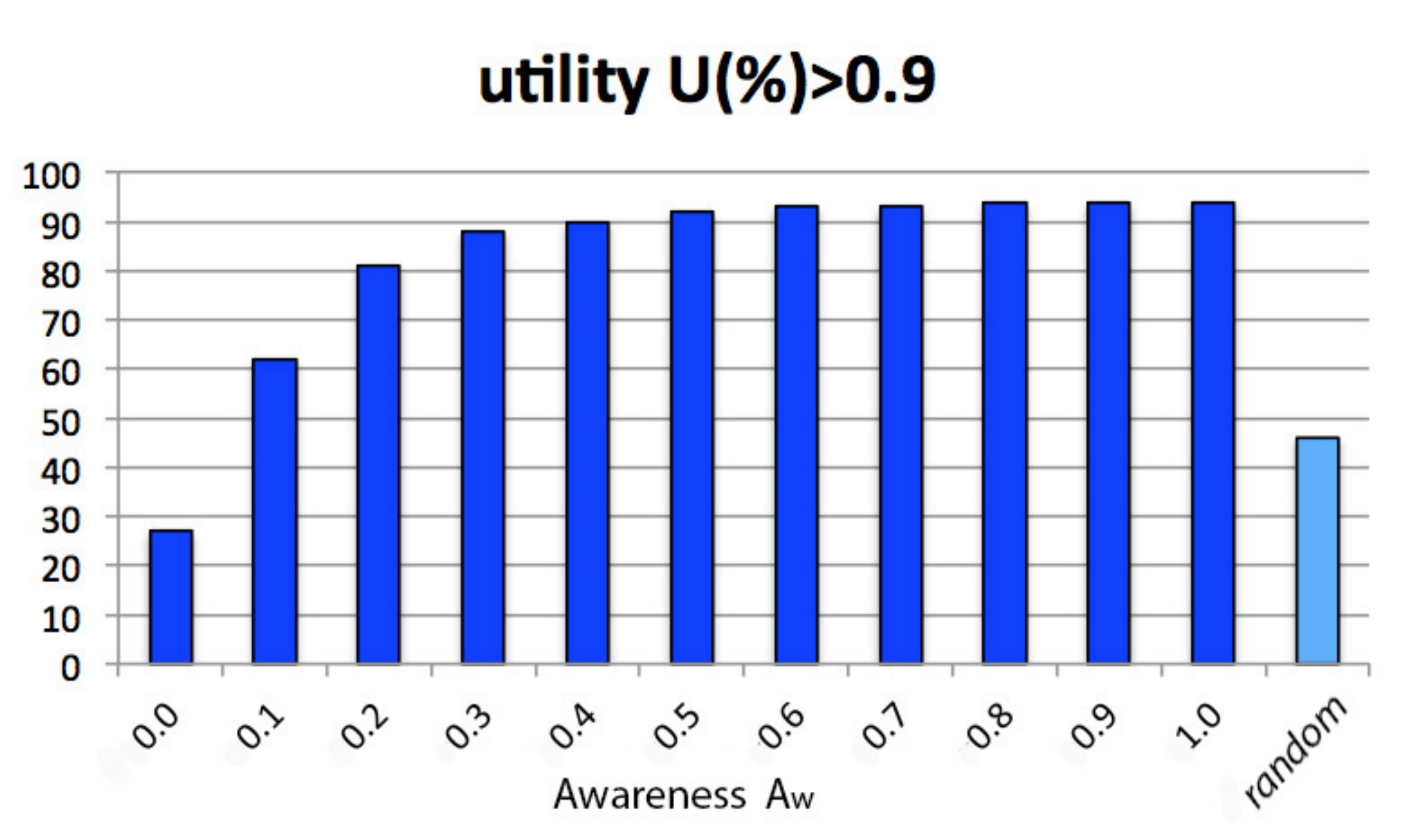}}
\caption{\footnotesize \emph{Market with ``target on product''.}
Behavior of $H(\%)>0.9$ and $U(\%)>0.9$ as a function of $\Aw$, for
$\Kn=1$ and $N_\Lev=20$. The case of a random consumer is also
considered.}
\end{figure}

We conclude the analysis of the case ``target on product'' by
performing two important tests. The first test aims at verifying the
influence of the attraction field for the various types of
consumers. In Table 3 we list the percentage of events (over the
total $N=25000$ events) in which each type of consumer ends her
journey within one of the seven clusters of the graph, ranked by
the decreasing values of their mass $M$ (these 
results are very robust, since the error is less than $1\%$). 
\begin{table}[h]
\begin{center}
\footnotesize
\begin{tabular}{|c|||c|c|c||c|c|c|c|c|c|c|}
\hline Consumer  & $\Kn$  & $\Aw$ & $N_\Lev$ & $M\!=\!33$ &
$M\!=\!26$ &
$M\!=\!24$ & $M\!=\!19$ & $M\!=\!17$ & $M\!=\!12$ & $M\!=\!8$\\
\hline \hline
\#1        &  $0.5$ &   $0$        &   4   &  $26$ &   $21$   &   $18$   &  $14$ &   $13$ &  $6$   & $3$   \\
\#2        &  $0.5$ &   $0$        &   20  &  $29$ &   $18$   &   $17$   &  $16$ &   $11$ &   $6$  & $3$ \\
\hline
\#3        &  $0.5$ &   $1$        &   4   &  $17$ &   $26$   &   $15$   &  $11$ &   $9$  &   $10$ & $13$    \\
\#4        &  $0.5$ &   $1$        &   20  &  $17$ &   $23$   &   $10$   &  $11$ &   $11$ &   $11$ & $16$    \\
\#5        &  $0.5$ & \emph{random}&   $-$ &  $19$ &   $22$   &   $22$   &  $13$ &   $10$ &   $9$  & $5$   \\
\hline \hline
\#6        &  $1$  &   $0$         &   4   &  $32$  &   $23$  &   $20$   &  $10$ &   $9$  &   $4$  & $2$    \\
\#7        &  $1$  &   $0$         &   20  &  $32$  &   $23$  &   $21$   &  $10$ &   $9$  &   $4$  & $2$ \\
\hline
\#8        &  $1$  &   $1$         &   4   &  $17$  &   $20$  &   $17$   &  $16$ &   $12$  & $11$  & $8$ \\
\#9        &  $1$  &   $1$         &   20  &  $22$  &   $20$  &   $17$   &  $13$ &   $11$  &  $9$  & $7$  \\
\#10       &  $1$  & \emph{random} &   $-$ &  $21$  &   $22$  &   $21$   &  $12$ &   $11$  &  $8$  & $5$   \\
\hline
\end{tabular}
\end{center}
\caption{\footnotesize \emph{Test for attractive field in a market
with ``target on product''.} The percentage of stops within each of
the seven clusters of the graph is reported as a function of their
mass $M$.}
\end{table}

\begin{figure}[h]
\centering
\includegraphics[width=0.5\textwidth]{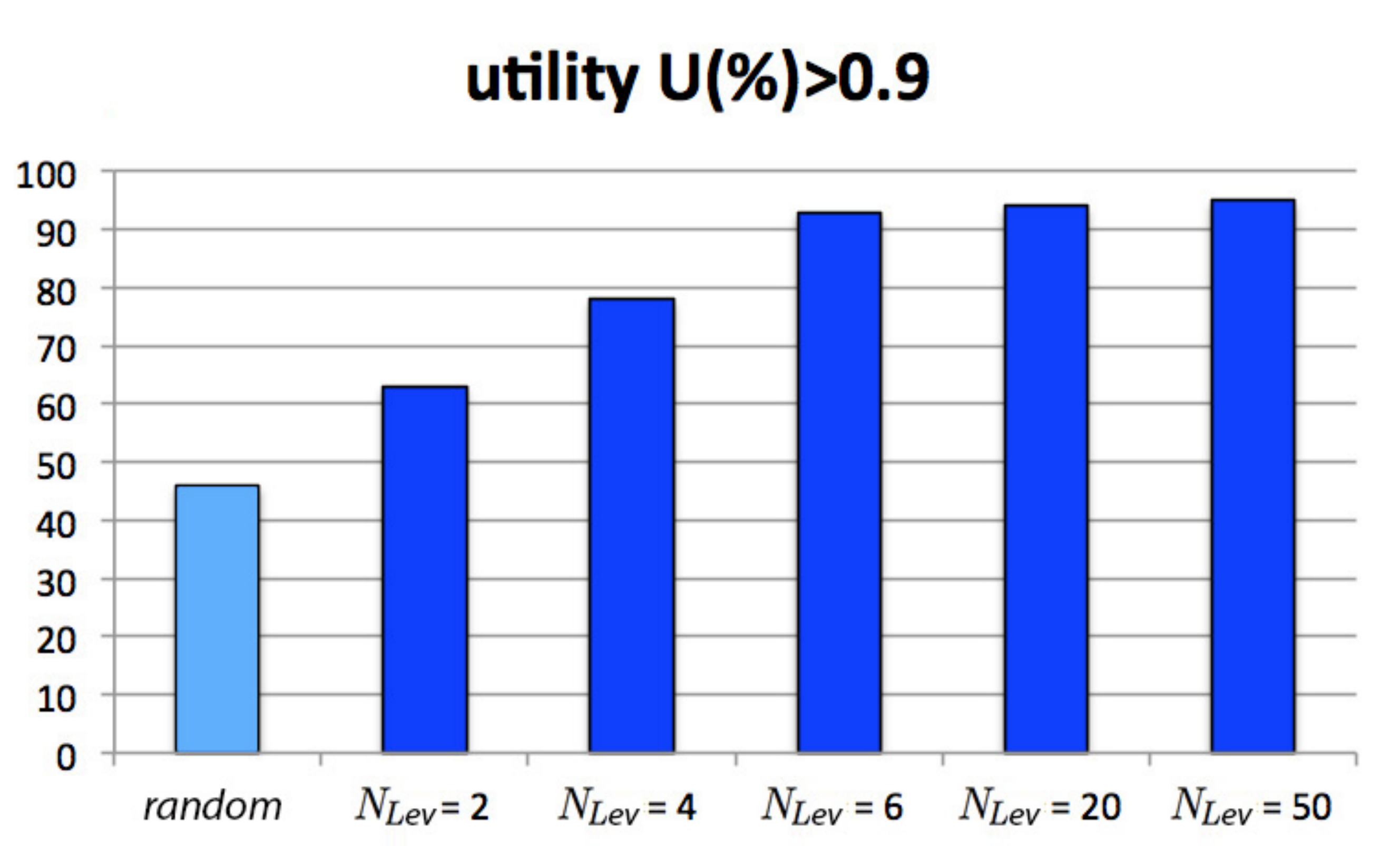}
\caption{\footnotesize \emph{Test for the role of indifference
levels in a market with ``target on product''.} The percentage of
events with final utility greater than $0.9$ is reported as a
function of the number of indifference levels.}
\end{figure}

As expected (see Section 2.4), only consumers characterized by
$\Aw=0$ are visibly attracted by clusters with high masses,
especially in case they are free of moving all over the graph (as
it happens for types $\#6$ and $\#7$, who possess total knowledge):
the percentage of stops is particularly high for the
cluster with $M=33$, and then it decreases along with decreasing
masses. For all the other types of consumers, including random ones,
the different number of stops in the various clusters is only due to
obvious statistical effects: in fact, clusters with a higher mass
also possess a larger number of nodes, hence the probability that a
buyer stops within them is proportionally higher, even with no
attraction effect whatsoever.

A second test concerns the effect of the number $N_\Lev$ of
indifference levels, the parameter determined on the basis of the
consumer's preference structure. In Figure 9 we report the
percentage of events (over the total $N=25000$) in which perfectly
informed consumers (type $\#9$, with $\Kn=1$ and $\Aw=1$) reach a
final utility greater than $0.9$ as a function of $N_\Lev$. The
analogous result for the random consumer (the same as in Figure
8(b)) is also reported for comparison. It is apparent, on one hand,
that the presence of only two indifference levels ($N_\Lev=2$) does
not help the informed consumer to perform much better than the
random one (only $17\%$ more). On the other hand, more than twenty
indifference levels ($N_\Lev>20$) do not appreciably improve her
performance. This explains why our simulations only take into
account the cases with $N_\Lev=4$ and $N_\Lev=20$.\footnote{The
previous tests are quite robust, and stay substantially unchanged
when the target does not coincide with an existing product.
Therefore we will not repeat them in the next section.}


\subsection{Target off product}

Here we summarize the results of simulations -- similar to those of
the previous section -- in the case that the target is on a point of
the two-dimensional metric space which does not coincide with any
terminal node of the graph. In this situation, even perfectly
aware consumers cannot reach the target, and their informative
journey is destined to end in a cul-de-sac (at time
$t_{\mathrm{end}}$). However, since nothing prevents that consumers
can approach the target quite early in their journey, we expect high
values of final utility for consumers with $\Aw=1$ and $\Kn=1$. On
the other hand, we also expect that, in the same conditions,
efficiency is not so high as in the target-on-product case: in fact,
for the target-off-product case, the journey goes on even after
the consumer has reached her minimum distance from the target at
time $t^*$, and so the ratio $t^*/t_{\mathrm{end}}$ might be quite
lower than $1$.

The distributions of both the efficiency and the final utility for
the 10 types of consumers is shown in Figure 10, where, as in Figure
7, in panel (a) we consider consumers with limited knowledge
($\Kn=0.5$), and in panel (b) we display consumers with complete
knowledge ($\Kn=1$). The obtained results essentially confirms our
expectations: group (b) performs better than group (a), in
particular for what concerns the final utility; further, perfectly
aware buyers (with $\Aw=1$) perform better than totally unaware ones
($\Aw=0$). It is worth noting, once again, the surprising effect of
randomness, namely, the good performance of random buyers with
respect to consumers with no awareness.

\begin{figure}
\centering \subfloat [][~~~Distributions for consumer types \#1-5
($\Kn=0.5$)]
{\includegraphics[width=0.485\textwidth]{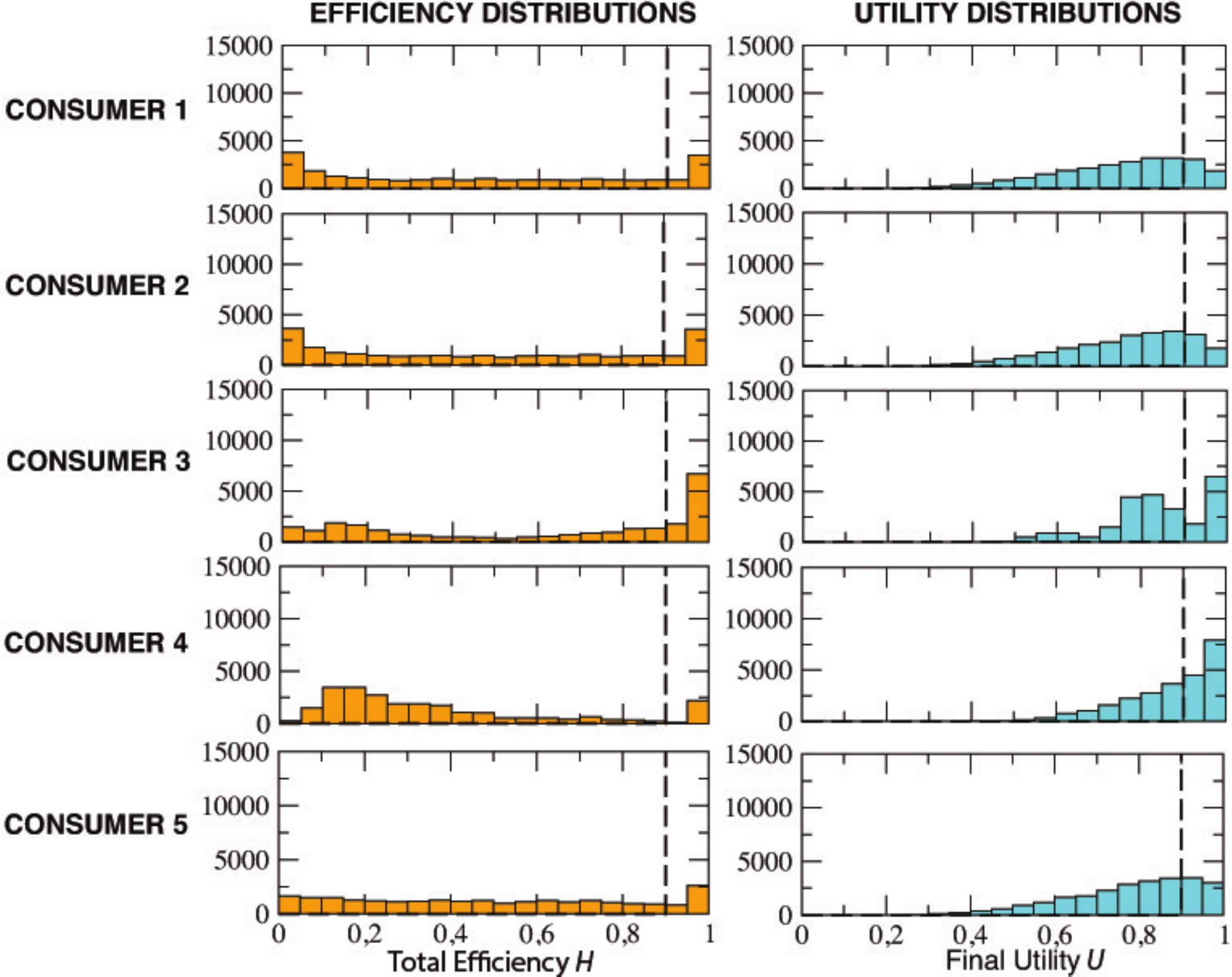}} \quad
\subfloat [][~~~Distributions for consumer types \#6-10 ($\Kn=1$)]
{\includegraphics[width=0.485\textwidth]{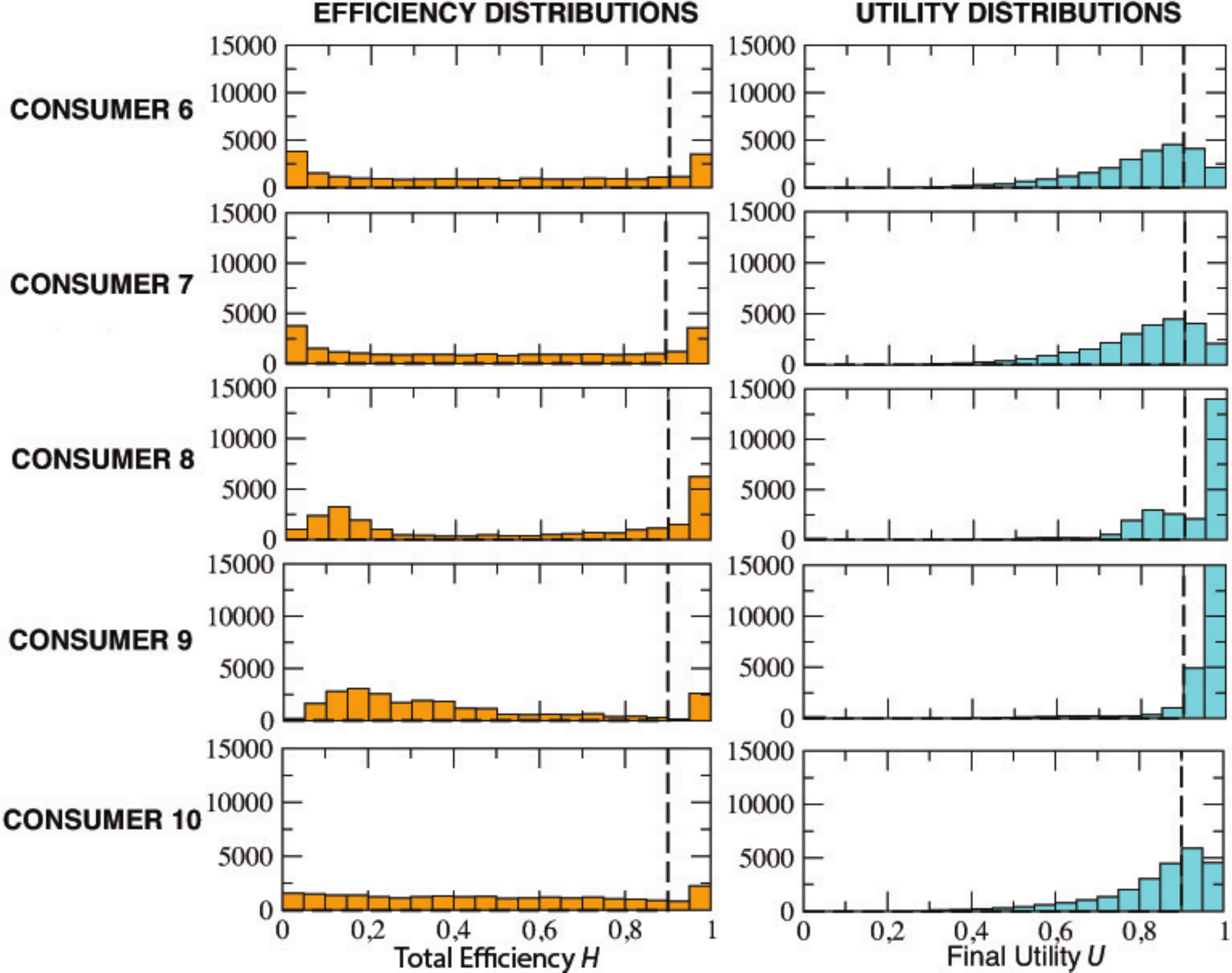}}
\caption{\footnotesize \emph{Market with ``target off product''.}
Distributions of efficiency and final utility calculated over
$N=25000$ simulation events for the 10 types of consumers. Dashed
lines indicate levels of efficiency and utility higher than 0.9.}
\end{figure}

These trends find further support in Table 4, where we report
details about the percentage of events with total efficiency $H>0.9$
and final utility $U>0.9$, along with social welfare and average
stopping time (and the corresponding standard deviations). In general --
comparing the results to those of Table 2 -- Table 4 shows lower
scores for all observed variables (except the stopping times, which are
higher) and all consumer types, thus indicating a global performance
that is (naturally) worse than the analogous one in the
target-on-product case.

\begin{table}[ht]
\begin{center}
\small
\begin{tabular}{|c|||c|c|c||c|c||c|c|c|c|}
\hline Consumer &  $\Kn$  &   $\Aw$   & $N_\Lev$  &  $H(\%)>0.9$ &
$U(\%)>0.9$ & $W$ &
$\sigma(W)$ & $T$ & $\sigma(T)$\\
\hline \hline
\#1        &  $0.5$ &   $0$   &   4   &  $18$ &   $19$   &   $0.75$    &  $0.16$ &   $49$ &   $25$   \\
\#2        &  $0.5$ &   $0$   &   20 &  $18$ &   $19$   &   $0.76$  &  $0.15$ &   $48$ &   $24$  \\
\hline
\#3        &  $0.5$ &   $1$   &   4   &  $34$ &   $33$   &   $0.84$    &  $0.12$ &   $35$ &   $15$    \\
\#4        &  $0.5$ &   $1$   &   20 &  $9$  &   $50$   &   $0.87$   &  $0.12$ &   $31$ &   $17$    \\
\hline
\#5        &  $0.5$ & \emph{random}    &   $-$ &  $14$ &   $26$   &   $0.78$  &  $0.15$ &   $46$ &   $25$   \\
\hline
\hline
\#6        &  $1$  &   $0$   &   4    &  $19$  &   $25$   &   $0.80$   &  $0.14$ &   $79$ &   $42$     \\
\#7        &  $1$  &   $0$   &   20  &  $19$  &   $25$   &   $0.80$  &  $0.14$ &   $79$  &   $42$   \\
\hline
\#8        &  $1$  &   $1$   &   4   &  $31$  &   $64$   &   $0.91$  &  $0.11$ &   $47$  &   $19$   \\
\#9        &  $1$  &   $1$   &   20  &  $11$  &   $89$   &   $0.94$  &  $0.10$ &   $33$  &   $19$   \\
\hline
\#10      &  $1$  &  \emph{random}  &   $-$&  $12$  &  $42$   &   $0.84$  &  $0.14$ &   $65$  &   $40$    \\
\hline
\end{tabular}
\end{center}
\caption{\footnotesize \emph{Market with ``target off product''.}
Global quantities are calculated over $N=25000$ simulation events
for the 10 types of consumers. We report, in order, the same
quantities as in Table 2: the results are very similar.}
\end{table}

\begin{figure}[ht]
\centering \subfloat [][Percentage of events with total efficiency
$H(\%)\!>\!0.9$]
{\includegraphics[width=0.485\textwidth]{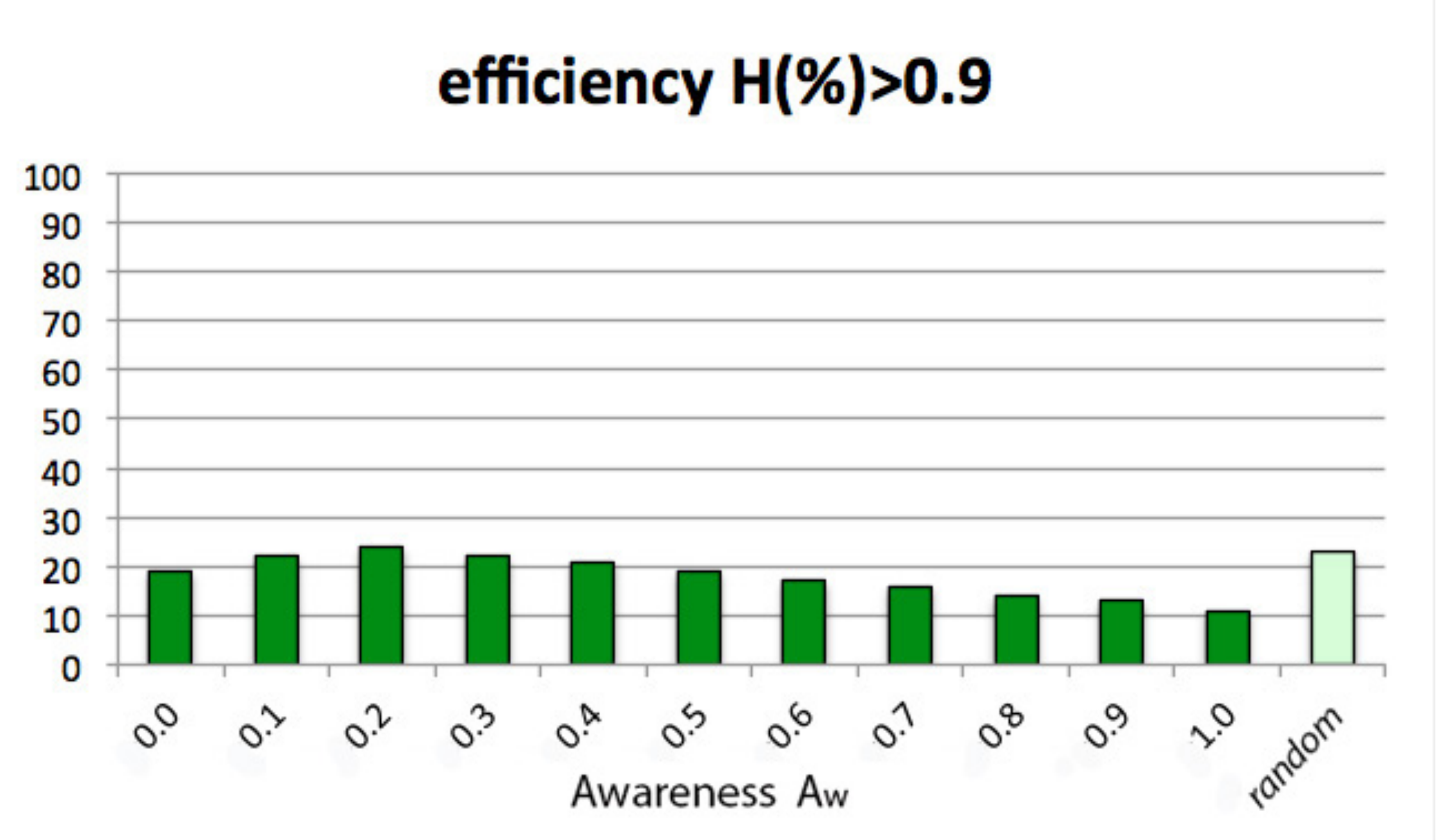}} \quad
\subfloat [][Percentage of events with final utility
$U(\%)\!>\!0.9$]
{\includegraphics[width=0.485\textwidth]{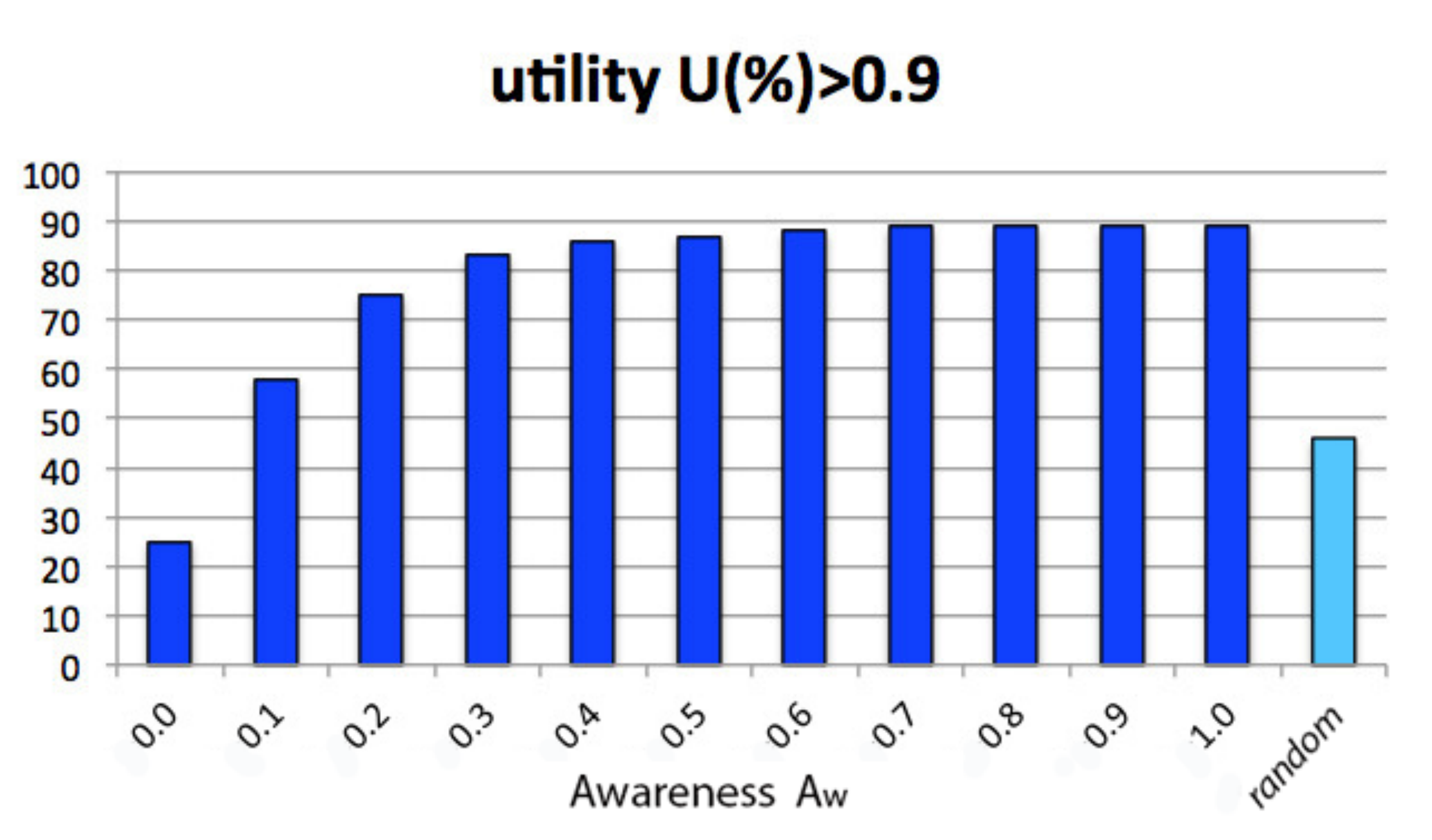}}
\caption{\footnotesize \emph{Market with ``target off product''.}
Behavior of $H(\%)>0.9$ and $U(\%)>0.9$ as a function of $\Aw$, for
consumers with $\Kn=1$  and $N_\Lev=20$. The case of a random consumer is also
considered.}
\end{figure}

In particular, comparing the performances of consumer $\#9$, which
is naturally the best type in both the tables, one notice that in
Table 4 the efficiency $H>0.9$ drops from $93\%$ down to $11\%$,
while the utility $U>0.9$ does not reach $90\%$ (against a $94\%$ of
Table 2); at the same time, the social welfare passes from $0.97$ to
$0.94$, and the average stopping time increases from $7$ to $33$ time
steps. On the other hand, comparing the random consumer $\#10$ in
the two tables, one can see that the collapse of performance is less
dramatic: this means that the random strategy confirms its superiority with respect to the completely uninformed one (visible by comparing consumers $\#10$ and $\#7$ in Table 4), and seems also to be unaffected by the position of the target (on or off product).

In Figure 11, the behavior of $H(\%)>0.9$ and $U(\%)>0.9$ (for
buyers with $\Kn=1$ and $N_\Lev=20$) is reported as a function of
consumer awareness, as already done in Figure 8. Compared with the
latter, it immediately appears that the final utility trend -- shown
in Figure 11(b) -- remains almost unchanged (it rapidly saturates to
a maximum value that is just slightly lower than in the
target-on-product case), but the shape of the efficiency behavior
looks now very different: indeed, due to the impossibility to reach
the target, the efficiency score never exceeds $30\%$, staying also
below the one of the random consumer for any value of awareness.
Thus, considering both efficiency and utility, the random consumer's
performance results again quite effective, not only with respect to
the totally unaware consumers, but also with respect those with
$\Aw>0$: again, this is particularly relevant in view of the fact
that gathering information has a cost, and so randomness appears
even more convenient than what one can derive from the results of
Figure 11(b).

Finally, let us to focus the attention on a strange effect visible
in both Tables 2 and 4. Some reader may have noticed that, when
consumers with $\Aw=1$ cannot reach the target (because either
$\Kn<1$ or the target is off product), the percentage of events
with efficiency $H$ greater than $0.9$ seems to {\it decrease} when
the discriminating ability of consumers increases (from $N_\Lev=4$
to $N_\Lev=20$), whereas (as expected) the final utility $U(\%)>0.9$
increases. This happens, in particular, for consumer types \#3 and
\#4 in Table 2, and types \#3, \#4, \#8 and \#9 in Table~4. Although
counterintuitive at a first sight, such a behavior has a natural
explanation if one deeper analyzes how the efficiency of a consumer
is affected by the presence of a small number of indifference levels
around the target (which is a symptom of a low discriminating
ability).
\begin{figure}[t]
\centering
\includegraphics[width=0.45\textwidth]{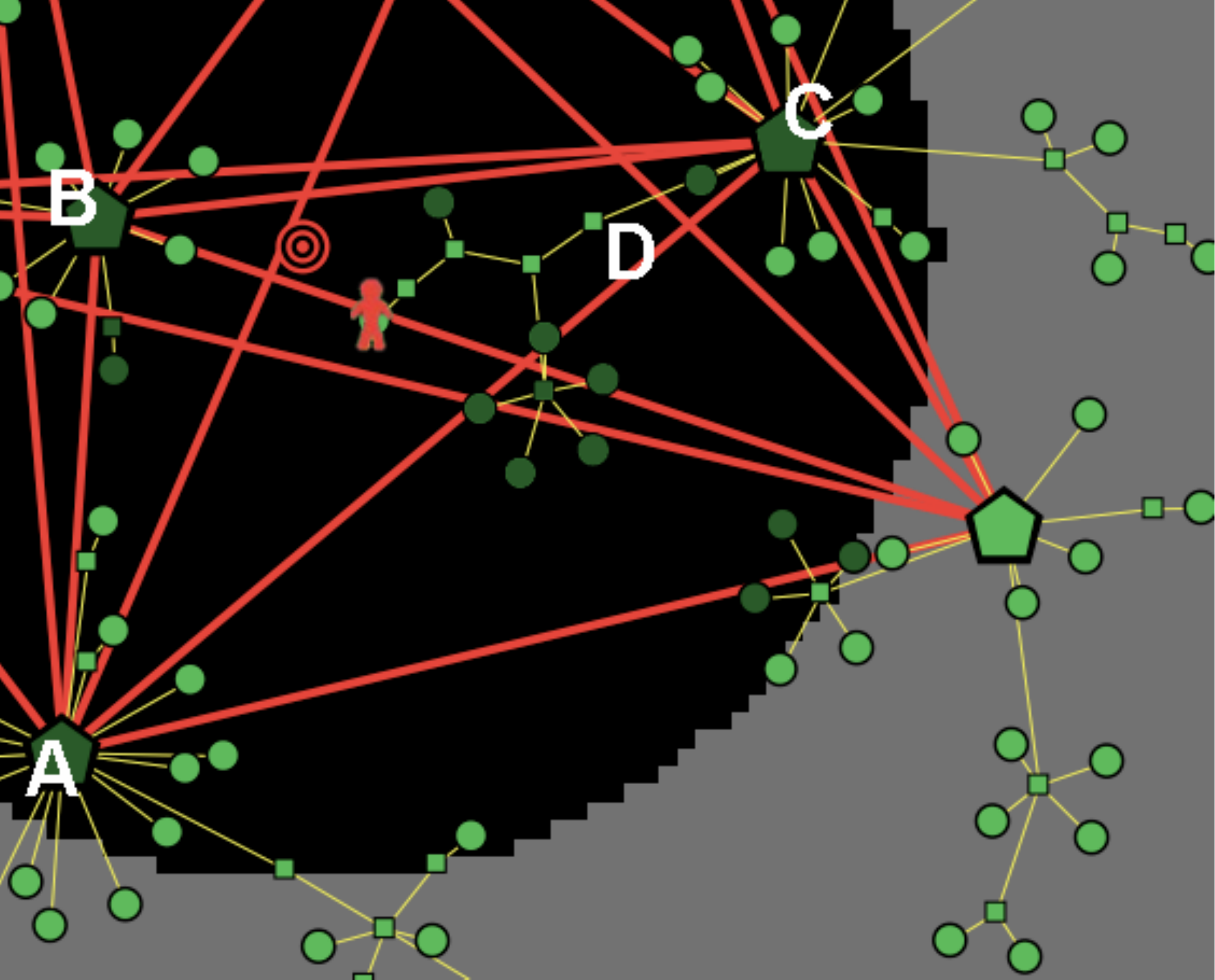}
\caption{\footnotesize \emph{Market with ``target off product'':
an example of the counterintuitive effect due to a small number of
indifference levels.} In an attempt to reach her ideal target, a
well aware and informed consumer with a low discriminating ability
(type $\#8$ of Table 4) remains trapped in a cul-de-sac, yet her
efficiency stays quite high.}
\end{figure}
In Figure 12 we show an example of informative journey for a
consumer with $\Kn=1$, $\Aw=1$ and $N_\Lev=4$, in the
target-off-product case (that is, we are considering consumer \#8
of Table 4). Due to her low discriminating ability, the buyer
considers all the nodes within the first indifference level (colored
in black) as equivalent: therefore, guided only by the degree of
nodes, she  wanders among the three hubs $A$, $B$ and $C$, with only
occasional raids down some branches of the respective clusters,
until she  finally ends up in the branch where is depicted in the
figure. As one can see, she  has reached (at, say, time $t^*$) the
product placed at the minimum possible distance from the target, and
so her satisfaction -- and, in turn, her utility -- jumps to its
(relative) maximum value (see Equations~(1) and~(2)). However, her
journey is bound to end soon: in fact, due to the high number of
previous visits, hub $C$ has been switched-off, hence the consumer's
fate is to remain trapped in a cul-de-sac. In particular, she  will
stop, after four more steps, on node $D$ at time
$t_{\mathrm{end}}=t^*+4$, thus producing an efficiency score very
close to $1$ (see Equation (3)); on the other hand, her utility
score may be quite lower than $1$. Translated into reality, this
effect captures the behavior of an undecided (and possibly well
informed) consumer who, due to her scarce discriminating ability,
oscillates for a while among different brands, and suddenly decides
to buy a certain product, even if it does not perfectly matches her
target.


\section{Conclusive Remarks}

In this paper we have presented a graph-based model of consumer
choice, which describes the hypothetical cognitive journey that each
individual experiences in the process of buying a product. The role
of the causes that influence the decision is measured by means of
behavioral differentiation in several parametric simulations. Our
results explicitly show the relevance of information and knowledge,
in the form of individual awareness, discriminating ability, and
perception of market structure.

We have focused our attention on some prototypical categories of
consumers, taking in account their subjectively distorted visions of
the market, as it seems to happen in everyday consumption
experiences. Many of our results confirm what one would naturally
expect: for example, a perfect knowledge of the market structure
paired with a high discrimination ability and a good individual
awareness usually determines a very satisfactory choice. On the
other hand, a few results of our simulations look rather surprising
-- and maybe intriguing -- in terms of individual satisfaction,
efficient strategies, and decision procedures, and therefore 
call for a deeper analysis of their explanation and consequences.

First of all, our model shows that consumers provided with a minimal
level of knowledge and information may unexpectedly reach very high
levels of utility. This appears to be in sharp contrast with the
classical paradigm that ``perfect information is mandatory to obtain
optimal results''. However, considered that actual markets are far
from being characterized by perfect information, our results provide
some justification for more realistic approaches, which do not rely
on perfect information as an unquestionable tenet of optimality.

Second, the results of our simulations consistently suggest that
whenever consumers fail to have a minimal level of knowledge and
information, random decisions will make them better off. This
translates into a new type of behavioral strategy, which may operate
as a sort of protective shield for the unaware category of
consumers. Said differently, a consumer, who wants to avoid that
social-economic forces -- advertising, bandwagon effects, persuasive
market power -- may well defeat market attraction by employing a random
approach, especially since the latter is free of charge. It is worth
noting that a random behavioral model could be a source of
inspiration for alternative strategies of both firms and policy
makers, for example concerning new anti-trust and competition laws.

A last consideration arising from the results of our simulations
concerns the emerging category of the ``informed-but-undecided''
consumer. In fact, for this peculiar type of buyers, it turns out
that the higher their discrimination ability is, the worse their
efficiency in consumption becomes. This seeming counterintuitive
effect has however an explanation: with a great capacity to distinguish 
differentiated characteristics of goods, the final efficiency can be very high 
just in case of few indifference levels.

Further research will be devoted to applying variations of our model
to different types of decision problems.

\end{document}